\documentclass[12pt,letterpaper]{article}
\usepackage[utf8]{inputenc}

\usepackage{graphicx,array}
\usepackage{url}
\usepackage{color}
\usepackage{latexsym}
\usepackage{amsthm}
\usepackage{amsmath}
\usepackage{amssymb}
\usepackage{amsfonts}
\usepackage[numbers,sort&compress]{natbib}
\usepackage{bm}
\usepackage{bbm}
\usepackage{slashed}
\usepackage{mathrsfs}
\usepackage{enumerate}
\usepackage{tikz}
\usepackage{siunitx}
\usepackage{mdframed}
\usepackage{setspace}  
\usepackage{esvect}
\usepackage{physics}
\usepackage{tikz-feynman}
\usepackage{enumitem}
\usepackage{tcolorbox}%

\usepackage{hyperref} 
\hypersetup{
    colorlinks=true,       
    linkcolor=red,          
    citecolor=blue,        
    filecolor=magenta,      
    urlcolor=blue           
}
\usepackage[all]{hypcap} 
\usepackage{multirow}
\usepackage{multicol}

\usepackage{natbib}
\usepackage[normalem]{ulem} 

\newcommand{\nc}{\newcommand}

\nc{\beq}{\begin{equation}}
\nc{\eeq}{\end{equation}}
\nc{\beqa}{\begin{eqnarray}}  
\nc{\eeqa}{\end{eqnarray}}  
\nc{\bit}{\begin{itemize}}  
\nc{\eit}{\end{itemize}}  

\newcommand{\eg}{{\it e.g.}}
\newcommand{\ie}{{\it i.e.}}

\usepackage{floatrow}
\newfloatcommand{capbtabbox}{table}[][\FBwidth]

\usepackage{blindtext}

\title{ 
 {\bf 
Flow-based Nonperturbative Simulation of First-order Phase Transitions 
 }
\author{\large Yang Bai and Ting-Kuo Chen}
\date{\small \it 
Department of Physics, University of Wisconsin-Madison, Madison, WI 53706, USA
}
}

\begin{document}

\maketitle

\setlength{\parskip}{0.2ex}

\begin{abstract}
We present a flow-based method for simulating and calculating nucleation rates of first-order phase transitions in scalar field theory on a lattice. Motivated by recent advancements in machine learning tools, particularly normalizing flows for lattice field theory, we propose the ``partitioning flow-based Markov chain Monte Carlo (PFMCMC) sampling" method to address two challenges encountered in normalizing flow applications for lattice field theory: the ``mode-collapse" and ``rare-event sampling" problems. Using a (2+1)-dimensional real scalar model as an example, we demonstrate the effectiveness of our PFMCMC method in modeling highly hierarchical order parameter probability distributions and simulating critical bubble configurations. These simulations are then used to facilitate the calculation of nucleation rates. We anticipate the application of this method to (3+1)-dimensional theories for studying realistic cosmological phase transitions.
\end{abstract}

\thispagestyle{empty}  
\newpage    
\setcounter{page}{1}  

\begingroup
\hypersetup{linkcolor=black,linktocpage}
\tableofcontents
\endgroup

\newpage

\section{Introduction}\label{sec:intro}

Cosmological first-order phase transitions (FOPTs) in the early universe have long been an intriguing field of study. With a wide range of extensions beyond the Standard Model (SM) of particle physics suggesting the possibility of cosmological FOPTs, there is a constant demand for rigorous predictions and measurements of the associated signatures. For instance, electroweak FOPTs are essential for explaining the matter-antimatter asymmetry through the mechanism of electroweak baryogenesis~\cite{Kuzmin:1985mm,Cohen:1993nk}. In addition, QCD could also undergo a FOPT with the assistance of new physics beyond the SM. For instance, late-collapsed domain walls with corresponding discrete symmetries anomalous under QCD can create a different thermal environment to induce QCD FOPTs~\cite{Bai:2023cqj}. Such transitions could potentially be probed by several pulsar timing array experiments aimed at detecting stochastic gravitational wave backgrounds (SGWBs)~\cite{NANOGrav:2020bcs,EPTA:2021crs,Goncharov:2021oub,Perera:2019sca}. Beyond the symmetries in the SM, FOPTs may occur in the dark matter sector, Grand Unified Models, quark and neutrino flavor models, with the produced gravitational waves serving as important messengers to probe early-universe physics (see Ref.~\cite{Athron:2023xlk} for a recent review).

The traditional method to study cosmological FOPT is based on the nucleation theory pioneered by Langer~\cite{Langer:1969bc} for non-relativistic classical systems, which was later generalized by Coleman~\cite{Coleman:1977py} and Linde~\cite{Linde:1980tt,Linde:1981zj} for zero- and finite-temperature relativistic quantum fields. The general picture is to identify the probability flux flowing from the metastable vacuum phase, through the saddle-point or critical-bubble configuration, to the stable vacuum phase as the nucleation rate. This picture allows one to factor the nucleation problem into a statistical (equilibrium) part and a dynamical (nonequilibrium) part and solve it using the method of saddle-point approximation. However, this semi-classical approach falls short of the following matters: 1) It cannot account for the fluctuations (and thus non-sphericity) of the bubbles in a realistic way. 2) The dynamical part requires the real-time evolution input of the system according to, \eg, the Langevin equation. 3)
In the case of strong-coupling models, such as QCD-type models, numerical nonperturbative tools are necessary for performing the calculations.
Nevertheless, as the nucleation picture given by Langer is still generally true for a system with hierarchical FOPT and particle scattering timescales, one can simply generalize the numerical method from the saddle-point approximation to the nonperturbative lattice field theory. For such studies of electroweak FOPTs, one can refer to Refs.~\cite{Moore:2000jw,Gould:2022ran}. More recently, Ref.~\cite{Gould:2024chm} has shown up to compare the nucleation rates calculated from the lattice formalism and the perturbative formalism for a real scalar theory in (3+1) dimensions. 

In the literature, lattice-based Markov chain Monte Carlo (MCMC) simulations of FOPTs continually encounter the following challenges:
\bit
\setlength\itemsep{0.1em}
\item[$\mathcal{A}$]: critical slowing down~\cite{Wolff:1989wq,DelDebbio:2004xh,Meyer:2006ty,Schaefer:2009xx,Schaefer:2010hu}; 
\item[$\mathcal{B}$]: mode collapse for multimodal distributions~\cite{Hackett:2021idh,Nicoli:2023qsl};
\item[$\mathcal{C}$]: rare-event sampling~\cite{gao2023rare}.
\eit
To tackle challenge $\mathcal{A}$, a recent method that utilizes normalizing flows (NFs), a class of generative machine learning models, has been demonstrated to be highly effective~\cite{Albergo:2019eim}. However, direct application of this flow-based method to FOPT simulations turns out to be ineffective because of challenges $\mathcal{B}$ and $\mathcal{C}$. On the other hand, the method of multicanonical MCMC (MMCMC)~\cite{Berg:1992qua} was proposed to address challenges $\mathcal{B}$ and $\mathcal{C}$. Even though MMCMC has been widely adopted in the nonperturbative  FOPT studies, it was noted in a recent paper, Ref.~\cite{Gould:2022ran}, that the development of more efficient algorithms is warranted. 

In this study, we propose a flow-based algorithm capable of simultaneously addressing all three challenges, which we term the ``partitioning flow-based (PF) MCMC" algorithm. The concept behind this algorithm is to train the NFs based on a reweighted target distribution (similar to the MMCMC algorithm), but instead of attempting to construct the complete distribution in one step, we ``slice" or ``partition" the phase space into multiple partitions, address them individually, and then ``glue'' them together. This method effectively tackles the three challenges as follows: 
\bit
\setlength\itemsep{0.1em}
\item[$\mathcal{A}$]: Flow-based MCMC algorithms inherently resolve the critical slowing down issue by decoupling the autocorrelation time of a Markov chain from the problem (or lattice) size~\cite{Albergo:2019eim} through the adoption of the independence Metropolis sampling algorithm~\cite{10.1214/aos/1176325750}.
\item[$\mathcal{B}$]: While conventional flow-based modeling struggles to capture the multimodal structure of the target distribution, the PFMCMC method treats it as several unimodal or monotonic modeling problems (except for regions around local minima, which are assumed to have a small enough hierarchy to be properly managed), thus enhancing stability and effectiveness.
\item[$\mathcal{C}$]: By partitioning the model into several sub-models, one can efficiently sample even from the rarest phase space regions using the corresponding sub-model.
\eit

To demonstrate the capability of the PFMCMC method, we first apply it to several toy 2D models. We show that it yields results consistent with those from the literature and outperforms certain revised flow-based algorithms. Additionally, we employ it in a realistic finite-temperature (2+1)D $\mathbb{Z}_2$-symmetric real scalar theory. Using the resulting probability distribution and bubble configurations, we illustrate a complete nucleation rate calculation for a benchmark scenario. Although our ultimate goal is to apply this method to a (3+1)D model for studying realistic cosmological FOPTs, the (2+1)D model we consider is highly motivated. For instance, in condensed matter physics, certain studies on the superfluidity of a 2D Bose gas rely on thermal mean field theory in (2+1) dimensions (see, for example, Refs.~\cite{Chien_2014,Konietin_2017,Pastukhov_2018,Pastukhov_2018_2}).

Our paper is organized as follows. In Section~\ref{sec:nucleation}, we review the classical nucleation theory outlined by Langer and its nonperturbative generalization based on lattice field theory. The semi-classical application of this theory to relativistic quantum fields is briefly reviewed in Appendix~\ref{sec:app-nucleation-QFT}. The flow-based algorithm for lattice field simulations is overviewed in Section~\ref{sec:method}, after which we discuss the challenges of bubble simulations and how the PFMCMC method can address these issues, demonstrated within a few toy 2D models. We introduce the finite-temperature (2+1)D $\mathbb{Z}_2$-symmetric real scalar theory, apply the PFMCMC method on one benchmark, and demonstrate a complete nucleation rate calculation based on the simulation results in Section~\ref{sec:sampling}, with the details of the scalar field theory given in Appendix~\ref{sec:app-finite-temp}. Finally, we discuss and conclude our study in Section~\ref{sec:conclusions}.

\section{Nucleation theory for first-order phase transitions}\label{sec:nucleation}

In this section, we first briefly review the classical nucleation theory outlined by Langer in Ref.~\cite{Langer:1969bc} to calculate the nucleation rates for the decays of a metastable or false vacuum. We then describe a convenient way to project the multi-dimensional functional space onto a one-dimensional order-parameter functional direction, following Ref.~\cite{Moore:2000jw}. After that, we introduce a method based on lattice field theory to calculate the nucleation rates nonperturbatively.  

\subsection{A review of classical nucleation theory}\label{subsec:nucleation:Langer}

For early-universe FOPTs, we are interested in the homogeneous nucleation or calculating the probability of a true-vacuum-phase bubble or droplet appearing in a system initially in the metastable vacuum phase. For a non-relativistic system, the theory of nucleation was developed by Langer through introducing a set of variables, $\eta_i$, to describe the collective degrees of freedom of the system~\cite{Langer:1969bc}. The distribution function or the probability density over the configurations and time, $\rho(\{\eta\}, t)$, is related to the free-energy of the system for equilibrium configurations (this is justified as long as the associated timescale for bubble nucleation is much longer than the particle thermalization timescale) by $\rho_{\rm eq}(\{\eta\}) \propto \mbox{exp}[- F(\{\eta \})/T]$. The phase transition begins from the system sitting at some initial point in the metastable region, later passing through the region of the saddle point, and then reaches the stable region. As pointed out by Langer, one could identify the probability flux through the saddle point with the nucleation rate without having to solve the complete time-dependent problem. 

Based on the saddle-point approximation, the nucleation rate can be expressed as the product of two factors: the dynamical factor $F_{\rm D}$ and the statistical factor $F_{\rm S}$. The statistical factor $F_{\rm S}$ can be calculated using the saddle-point approximation in Ref.~\cite{Langer:1969bc}, while the dynamical factor $F_{\rm D}$ is the exponential growth rate of a bubble that just exceeds the critical size, and is proportional to the negative eigenvalue of the free energy second-order differential matrix in $\eta_i$, evaluated at the saddle point. The dynamical factor was calculated in Ref.~\cite{Langer:1969bc} for a non-relativistic system and in Ref.~\cite{Csernai:1992tj} for a relativistic system. 

For a relativistic quantum field theory, which is more relevant when describing early-universe FOPTs,  Coleman has developed the formula for zero-temperature theories~\cite{Coleman:1977py} which was later generalized for finite-temperature theories by Linde~\cite{Linde:1980tt,Linde:1981zj} (see Appendix~\ref{sec:app-nucleation-QFT} for details). For our numerical lattice simulation of a simple scalar field model in $(2+1)$ dimensions, the general picture depicted by Langer is still applicable, although the lattice simulation of the phase transition does not need to rely on the saddle-point approximation. 

\begin{figure}[t!]
    \label{fig:transition-cartoon}
    \centering $\vcenter{\hbox{\includegraphics[width=0.46\textwidth]{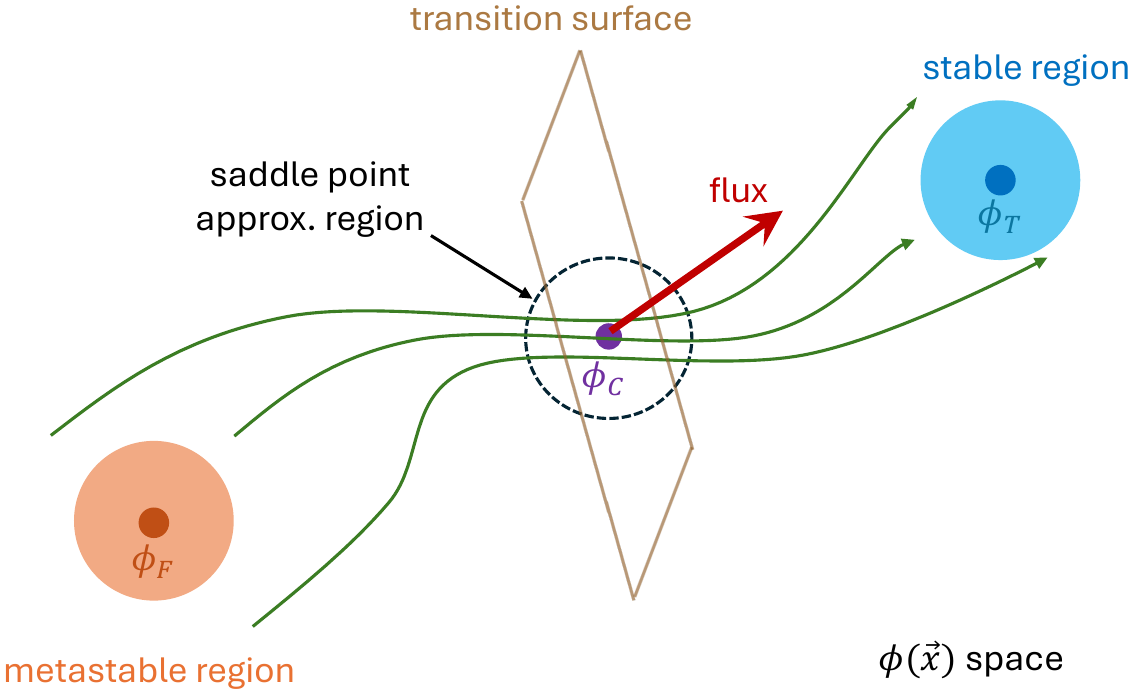}}}$ \hspace{5mm}   $\vcenter{\hbox{\includegraphics[width=0.46\textwidth]{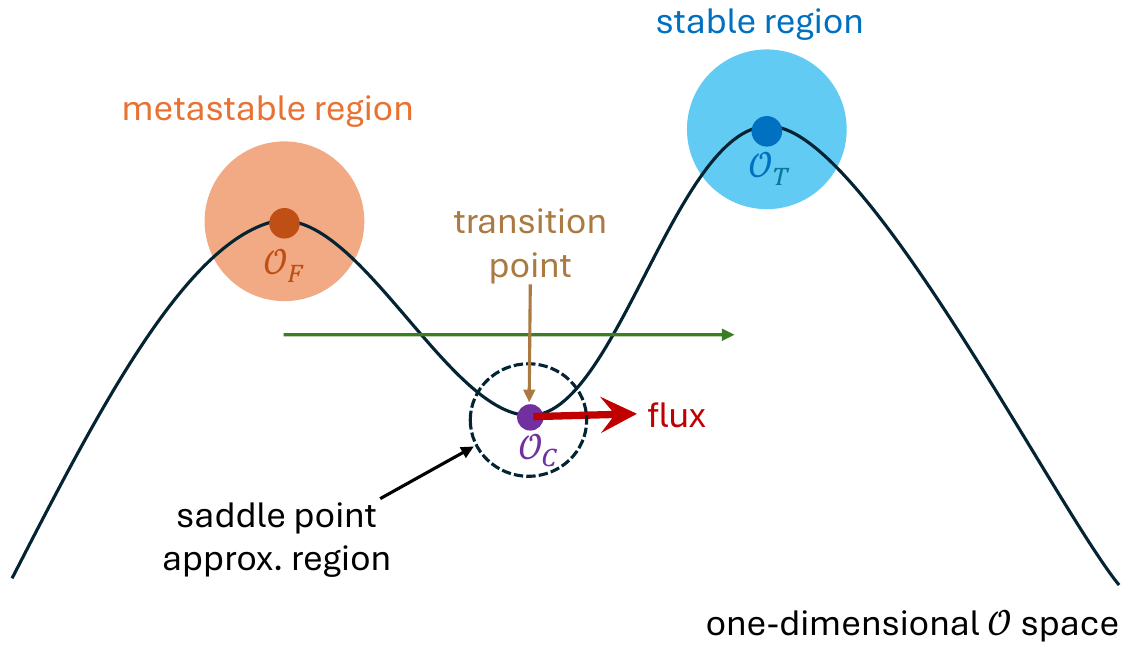}}}$
    \caption{{\it Left:} A cartoon illustration of the flow of field configurations from the metastable region to the stable region through the transition surface. Here, $\phi_F$ and $\phi_T$ represent the field configurations in the metastable (false) phase and the stable (true) phase, respectively. $\phi_C$ denotes the critical-bubble configuration, with ``flux" indicating the flow direction of the field configuration at $\phi_C$, and ``transition surface" referring to the hyper-dimensional surface perpendicular to the flux direction.
    {\it Right:} A similar illustration to the left panel, but with the hyper-dimensional functional space projected onto the one-dimensional measurable $\mathcal{O}$ space. The relative heights of the metastable, stable, and transition-point regions indicate their different statistical probabilities. Depending on the choice of the functional $\mathcal{O}[\phi(\vec{x})]$, the locations of $\mathcal{O}_{F, C, T}$ may be close to but do not exactly match to the configurations $\phi_{F, C, T}$ in the left panel.    
    }
\end{figure}

In the left panel of Fig.~\ref{fig:transition-cartoon}, we show a cartoon picture that illustrates various paths along which the configurations $\{\phi(\vec{x})\}$ flow from the metastable region to the stable region. Between these two regions, field configurations are supposed to pass through the critical-bubble region (for simplicity, only one critical-bubble configuration is assumed to exist). The phase space of the critical-bubble configuration or the saddle-point approximation region is smaller than those of the metastable and stable regions because the critical bubble sits at a local maximum of the free energy spectrum at least in one direction of the configuration space. A field configuration sitting at a point slightly above the critical bubble is expected to flow along the direction that reduces the free energy, or the ``flux" direction, in Fig.~\ref{fig:transition-cartoon}. Perpendicular to the flux direction, one could define the transition surface that is co-dimension one in the configuration space. The nucleation rate could be obtained by calculating both the statistical factor (the rareness of the critical-bubble configuration) and the dynamical factor (the quickness of the probability passing the critical-bubble point) around the saddle-point approximation region.  

In the right panel of Fig.~\ref{fig:transition-cartoon}, we show the projected one-dimensional subspace of $\{\phi(\vec{x})\}$ defined through a ``measurable" operator $\mathcal{O}[\phi(\vec{x})]$ that is a functional of the field configurations. The motivation for introducing such a measurable is for the convenience of later numerical simulations of FOPT. The independence of nucleation on a particular choice of $\mathcal{O}$ is demonstrated in Ref.~\cite{Moore:2000jw} (see also Ref.~\cite{Gould:2024chm} for the explorations of numerically friendly measurables). For a suitable choice of $\mathcal{O}$, there are still three regions: the metastable region, the transition point, and the stable region. The flow of field configurations in the left panel becomes a simple straight flow from the metastable region (left) to the stable region (right) that passes through the transition point in between. The relative heights of different regions indicate their different statistical probabilities. For a linear projection from $\{\phi(\vec{x})\}$ to $\mathcal{O}$, the locations of $\phi_{F,C,T}$ in the left panel should match to those of $\mathcal{O}_{F,C,T}$ in the right panel. This is not generally true for a non-linear projection. However, for a suitable choice of $\mathcal{O}$, the three regions on the left panel should have large overlaps with the corresponding regions on the right panel. Therefore, one can focus on the transition point region around $\mathcal{O}_C$ to calculate the nucleation rate. 

Based on the probability distribution $P(\mathcal{O})$ of the measurable, one can define a ``working region", $[\mathcal{O}_c - \epsilon/2, \mathcal{O}_c + \epsilon/2]$ with a small $\epsilon$, around the saddle-point approximation region (the numerical calculation can extend beyond the saddle-point approximation) to calculate the nucleation rate.  First, the statistical factor as defined in the Langer's method becomes
\beqa
\label{eq:F_S}
F_{\rm S} = \frac{P([\mathcal{O}_c - \epsilon/2, \mathcal{O}_c + \epsilon/2])}{\epsilon\, P(\mathcal{O} < \mathcal{O}_c - \epsilon/2)} ~,
\eeqa
which is the ratio of the probability density containing the critical bubble over the probability of the metastable phase configurations.

Second, the dynamical factor is proportional to the probability flux, $|d\mathcal{O}/dt|_{\mathcal{O}_c}$, at the transition point or the ``critical separatrix". In a practical numerical simulation, however, one needs to multiply the flux by a ``success factor", $\mathrm{S}$, to account for two facts: 1) Not every field evolution trajectory that passes through the critical point will eventually reach the stable region. 2) Some trajectories may have multiple turnarounds around the critical point before reaching the stable region, but they only overall contribute to one tunneling. This factor depends on either even or odd number of times that the trajectory crosses the critical point, $N_{\rm crossings}$, and is given by $\mathrm{S} = (N_{\rm crossings}\,\mbox{mod}\,2)/N_{\rm crossings}$. As one does not distinguish between the crossing in one direction and the opposite during the simulation, a factor of $\frac{1}{2}$ should be further multiplied in order to only account for metastable $\rightarrow$ stable trajectories. Consequently, the dynamical factor is~\cite{Moore:2000jw} 
\beqa
\label{eq:F_D}
F_{\rm D} = \left\langle \frac{1}{2}\,\mathrm{S}\, \Big|\frac{d\mathcal{O}}{dt}\Big|_{\mathcal{O}_c} \right\rangle = \left\langle \frac{1}{2}\,\frac{(N_{\rm crossings}\,\mbox{mod}\,2)}{N_{\rm crossings}}\,\Big|\frac{d\mathcal{O}}{dt}\Big|_{\mathcal{O}_c} \right\rangle ~,
\eeqa
where $\langle \rangle$ means averaging over numerical samples. 

Combining the two factors in Eqs.~\eqref{eq:F_S} and \eqref{eq:F_D}, one has the homogeneous nucleation rate per volume as
\beqa
\label{eq:rate-full}
\Gamma = \frac{F_{\rm D} \, F_{\rm S}}{V} = \left\langle \frac{1}{2}\,\frac{(N_{\rm crossings}\,\mbox{mod}\,2)}{N_{\rm crossings}}\,\Big|\frac{d\mathcal{O}}{dt}\Big|_{\mathcal{O}_c} \right\rangle \,\times \, \frac{P([\mathcal{O}_c - \epsilon/2, \mathcal{O}_c + \epsilon/2])}{\epsilon\, P(\mathcal{O} < \mathcal{O}_c - \epsilon/2)\,V} ~,
\eeqa
where $V$ stands for a large enough volume that contains the critical-bubble configuration. We remark that the exact value of $\epsilon$ does not affect the overall result much as long as it is reasonably small. There are counteracting dependencies on $\epsilon$ for the statistical and dynamical factors which eventually cancel out. The details of the argument can be found in Ref.~\cite{Moore:2000jw}.

\subsection{Nonperturbative calculations based on lattice field theory}\label{subsec:nucleation:lattice}

Using lattice field theory as a nonperturbative method to study FOPTs has a few advantages: 1) One is not restricted to perturbative theories, but could also study theories with nonperturbative couplings such as QCD-type models. 2) The calculations for nucleation rates or other quantities can go beyond the saddle-point approximation as in Langer's method. 3) The field configurations sampled numerically could be used to study other physics related to the FOPT, including the properties of non-spherical critical bubbles and gravitational wave productions. In this work, we focus on calculating the bubble nucleation rates of FOPTs, which is one of the most important quantities pertaining to the history of cosmological phase transitions. Our following lattice field theory setup follows closely to the general strategy in Ref.~\cite{Moore:2000jw}, where the detailed numerical implementations for various terms in Eq.~\eqref{eq:rate-full} have been addressed. Some recent studies following Ref.~\cite{Moore:2000jw} can be found in Refs.~\cite{Moore:2001vf,Gould:2022ran,Gould:2024chm}. Our main goal is to incorporate modern machine-learning tools into the studies of FOPTs.  

The calculation for the dynamical factor in Eq.~\eqref{eq:F_D} is fairly straightforward at least for a simple model with only a few fields. One could make a large sample of field configurations around the transition point (see the $\mathcal{O}_c$ region in the right panel of Fig.~\ref{fig:transition-cartoon}) and use the real-time Langevin equation to evolve the configurations following~\cite{Goldenfeld:1992qy,Moore:2000mx,critical-pheno}
\begin{equation}\label{eq:Langevin}
    \frac{d\phi}{dt} = -\frac{\delta S[\phi]}{\delta\phi} + \nu(t, \vec{x}) ~.
\end{equation}
Here, $S[\phi]=T\times S_{\rm E}[\phi]$ with $S_{\rm E}[\phi]$ being the temperature-reduced Euclidean action for studies of thermal FOPT [see Eq.~\eqref{eq:action} for the example of a scalar field theory]; $\nu(t,x)$ is a Gaussian white noise satisfying 
\begin{equation}
    \big\langle\nu(t,\vec{x})\big\rangle_\nu = 0 ,\qquad \big\langle\nu(t,\vec{x})\,\nu(t^\prime,\vec{x}^\prime)\big\rangle_\nu=2\,T\,\delta^{(d-1)}(\vec{x}-\vec{x}^\prime)\,\delta(t-t^\prime) ~,
\end{equation}
with $\langle \rangle_\nu$ representing the ensemble average over $\nu$ samples and $d-1$ as the number of spacial dimensions, which follows a Gaussian distribution with a width of $\sqrt{2\,T}$, \ie,
\beqa
    d\rho[\nu(t, \vec{x})] = d\nu(t, \vec{x})\,e^{-\int d^{d-1}x\,dt\,\nu^2(t,\vec{x})/(4\,T)} ~.
\eeqa
For the probability flux, one starts with a field configuration in the working region $[\mathcal{O}_c - \epsilon/2, \mathcal{O}_c + \epsilon/2]$ and evolves it both backward and forward in time according to Eq.~\eqref{eq:Langevin}. 
One can define the probability flux $\vert d\mathcal{O}/d t\vert_{\mathcal{O}_c}\approx\vert \Delta\mathcal{O}/\Delta t\vert_{\mathcal{O}_c}$ for a small time interval $\Delta t$, which does not distinguish the metastable $\rightarrow$ stable direction from the opposite direction.
The success factor $\mathrm{S}$ in Eq.~\eqref{eq:F_S} depends on whether the evolution trajectory eventually ends up in the opposite region (metastable vs. stable). If not, $N_{\rm crossings}$ is an even number and $(N_{\rm crossings}\, \mbox{mod}\,2) = 0$, indicating that this trajectory does not contribute to the dynamical factor. On the contrary, $(N_{\rm crossings}\, \mbox{mod}\,2) = 1$ for an odd $N_{\rm crossings}$, and this trajectory will contribute to the dynamical factor (see Fig.~\ref{fig:flux} for two examples of the trajectories from our numerical simulations). Finally, a factor of $1/N_{\rm crossings}$ is multiplied to account for the fact that there is only one effective crossing, as mentioned previously.

The calculation of the statistical factor in Eq.~\eqref{eq:F_S} requires a large sample of field configurations to obtain a reliable probability distribution, $\rho_{\rm eq}(\{\phi\}) \propto \mbox{exp}[- F(\{\phi \})/T]$ (see the right panel of Fig.~\ref{fig:transition-cartoon} in the one-dimensional measurable space). In the context of cosmological phase transition, the critical-bubble configuration with $\mathcal{O}_c$ could be highly suppressed by $\mathcal{O}(e^{-100})$ as a result of the hierarchy between the phase transition temperature and the Hubble scale. On top of this huge suppression factor, the multimodal structure of the probability distributions (see the right panel of Fig.~\ref{fig:transition-cartoon}) also makes it difficult for traditional MCMC methods to model them properly due to their ``mode-seeking'' behavior~\cite{mode-seeking}. Furthermore, traditional MCMC methods usually suffer from the critical slowing down issue that imposes a barrier to scalability. In lattice field theory, this issue is usually manifest in the lattice-size-dependence of the autocorrelation time of a constructed Markov chain~\cite{Albergo:2019eim}. For convenience, we have named the three challenges that traditional MCMC methods are usually faced with when calculating the probability distribution of measurable values 
as: $\mathcal{A}$: critical slowing down; $\mathcal{B}$: mode collapse for multimodal distributions; $\mathcal{C}$: rare-event sampling.  

Before we discuss our algorithm that adopts modern machine learning to overcome the three challenges, we also briefly comment on the existing method of MMCMC (see also Sec.~\ref{subsec:method:challenges}) introduced in Ref.~\cite{Berg:1992qua} where a weight function in terms of the measurable is introduced to flatten the target probability distribution for more efficient samplings. After the samplings are finished, the results are then reweighted back to retrieve the correct statistics. This method requires some initial (``well educated") guess for the right weight function and was already used and commented in Ref.~\cite{Moore:2000jw} (see Ref.~\cite{Gould:2022ran} for a recent application and request for the development of more efficient algorithms).

\section{A flow-based simulation algorithm}\label{sec:method}

In this section, we develop a flow-based algorithm to address all three challenges mentioned previously and outline the procedure to calculate the nucleation rates of FOPTs. For challenge $\mathcal{A}$, it has been shown in the literature that an independence Metropolis algorithm based on NFs naturally solves the critical slowing down issue~\cite{Kobyzev_2021, Albergo:2019eim}. As a family of generative models, NFs can model the intricate probability distributions by transforming a simple base distribution into a complex distribution via a series of invertible and differentiable transformations. Usually, these transformations are implemented with trainable deep neural networks. When applied to an independence Metropolis sampler (see Sec.~\ref{subsec:method:flow}), the autocorrelation time of the constructed Markov chain does not scale with the lattice size but only depends on how closely the trained NF model matches to the target distribution~\cite{Albergo:2019eim}. However, the simple application of NFs to lattice field theory does not solve challenge $\mathcal{B}$ (the mode-collapse problem), which has been discussed in Refs.~\cite{Hackett:2021idh,Nicoli:2023qsl}, or challenge $\mathcal{C}$ (the rare-event sampling problem), which has been discussed in Ref.~\cite{gao2023rare}. We will discuss these issues in detail and propose a partitioning flow-based method to overcome these challenges and apply it to bubble nucleation rate calculations. 

\subsection{Flow-based sampling in lattice field theory}\label{subsec:method:flow}

In lattice field theory, one uses MCMC approaches to generate field configurations according to the target distribution
\begin{equation}
    p(\phi) = e^{-S_{\rm E}[\phi]}/Z ,~ \quad \mbox{with}\quad Z=\int\prod_{j=1}^Dd\phi_j\,e^{-S_{\rm E}[\phi]} ~,
\end{equation}
where $j$ indexes the $D$ components of $\phi\in\mathbb{R}^D$ and $\phi$ is defined to be a vector with $D$ real components 
 matching the spacetime degrees of freedom for a real scalar $\phi(\vec{x})$ evaluated on a finite, discrete spacetime lattice. Within the flow-based framework, one constructs a Markov chain according to $p(\phi)$ via the independence Metropolis sampler~\cite{10.1214/aos/1176325750}. For each step $i$ of the chain, an updated proposal $\phi^\prime$ is generated by sampling from $\tilde{p}(\phi)$ [a function to be optimized to approximately match the true $p(\phi)$], independent of the previous configuration $\phi^{(i-1)}$. The step is accepted with probability
\begin{equation}
    A(\phi^{(i-1)},\phi^\prime) = {\rm min}\left[1, \frac{\tilde{p}(\phi^{(i-1)})}{p(\phi^{(i-1)})}\frac{p(\phi^\prime)}{\tilde{p}(\phi^\prime)}\right] ~.
\end{equation}
If the proposal is accepted, $\phi^{(i)}=\phi^\prime$, otherwise $\phi^{(i)}=\phi^{(i-1)}$.

\begin{figure}[th!]
    \label{fig:NF-cartoon}
    \centering $\vcenter{\hbox{\includegraphics[width=0.8\textwidth]{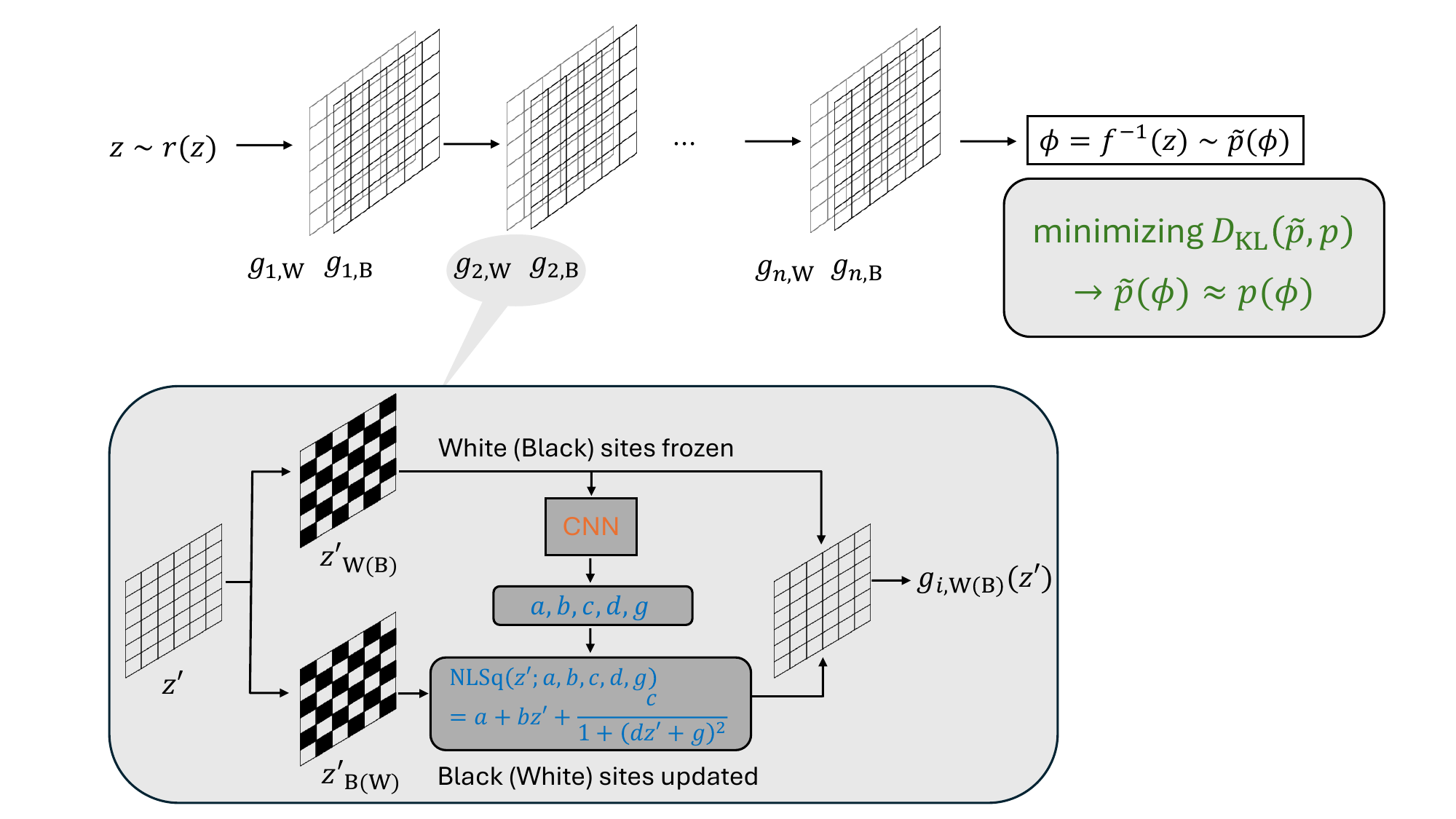}}}$ 
    \caption{
    A schematic plot of the flow model used in this study. In each iteration, a set of random variables on the lattice $z$ (a vector of $\mathbb{R}^D$) is drawn from the prior distribution $r(z)$. When the random variables pass through each (bi-)layer of the flow model, the white and black lattice sites are updated according to $g_{i,{\rm W}}$ and $g_{i, {\rm B}}$ sequentially. In each $g_{i,{\rm W}({\rm B})}$, the white(black) sites are frozen, and black(white) sites are updated according to the non-linear squared (NLSq) function. The parameters,  $a,b,c,d,g$, of the NLSq function are to be optimized and are parametrized with a convolutional neural network (CNN) using the inputs from the frozen sites. 
    The output is then forwarded to the next (bi-)layer until the last layer is reached. The final outputs are the transformed random variables $\phi=f^{-1}(z)\sim \tilde{p}(\phi)$. By minimizing the loss function $D_{\rm KL}(\tilde{p}\Vert p)$, the model is trained to sample $\phi$ from $\tilde{p}(\phi)\approx p(\phi)$.
    Our particular model structure has roughly 500K trainable hyperparameters.
    }
\end{figure}

To construct $\tilde{p}(\phi)$, we construct a NF $f$ that maps from $z\sim r(z)$ to $\phi=f^{-1}(z)\sim\tilde{p}(\phi)$, where from using Jacobian one has 
\begin{equation}
    \tilde{p}(\phi) = r\big(f(\phi)\big)\,\left\vert\det\frac{\partial f(\phi)}{\partial\phi}\right\vert ~.
\end{equation}
Here, the prior distribution function $r(z)$ with $z\in\mathbb{R}^D$ could be some simple known function (Gaussian distribution for instance). The functional mapping from $z\xrightarrow[]{f^{-1}} \phi$ or the inverted one $\phi\xrightarrow[]{f} z$ could be very complicated. One could construct $f$ by the composition of {\it coupling layers} and choose a triangular matrix for the transformation of each layer such that the determinant of the Jacobian can be efficiently computed~\cite{dinh2017density}. A schematic illustration of the flow model used in our study is shown in Fig.~\ref{fig:NF-cartoon}.

The model is optimized through minimizing the (approximate) shifted reverse Kullback-Leibler (KL) divergence~\cite{KL-divergence},
\begin{equation}\label{eq:reverse-KL}
\begin{aligned}
    L(\tilde{p},p) &\equiv D_{\rm KL}(\tilde{p}\Vert p)-\log Z \\
    &= \int\mathcal{D}\phi\,\tilde{p}(\phi)\,\big[\log\tilde{p}(\phi)+S_{\rm E}[\phi]\big]
    \approx \frac{1}{N}\sum_{\phi\in{\rm batch}}\big[\log\tilde{p}(\phi)+S_{\rm E}[\phi]\big]
    ~.
\end{aligned}
\end{equation}
Physically, shifting the loss function by the constant $Z$, which is in general intractable, does not affect the simulation outcome, thus justifying the choice of the loss function. In the last step, the loss is calculated by using a batch of $N$ samples at each epoch from the model $\{\phi\sim\tilde{p}\}$.
As proven and experimented in Ref.~\cite{Albergo:2019eim}, this flow-based MCMC sampler is immune to the critical slowing down problem, and the integrated autocorrelation time is only determined by the acceptance rate of the Markov chain, which is in turn determined by how well $\tilde{p}(\phi)$ assimilates $p(\phi)$. Moreover, training with the reverse KL divergence does not require a pre-generated dataset, which is another appealing feature of this method.

\subsection{Mode-collapse and rare-event sampling}\label{subsec:method:challenges}

One major challenge for this flow-based method is the ``mode-seeking'' behavior of the reverse KL self-training~\cite{mode-seeking,DelDebbio:2021qwf} that will possibly lead to the ``mode-collapse'' problem (one mode with a higher probability is identified while the additional lower-probability modes are ignored or underestimated). The dynamics of reverse KL training cause initial overconcentration on peaks, which has been observed to manifest as a tendency for the sampler to initially collapse onto a subset of modes. If the model never samples from a given mode, it is not penalized for badly mismodeling that excluded mode, since we cannot detect whether a mode has been missed as the normalizing factor $Z$ is unknown based on the loss function defined in Eq.~\eqref{eq:reverse-KL}. Ref.~\cite{Hackett:2021idh} has proposed several methods to mitigate the problem, and Ref.~\cite{Nicoli:2023qsl} has further compared the scalabilities of reverse and forward KL training schemes~\footnote{These refer to the training using the reverse and forward (or normal) KL divergence, respectively. For the forward KL, the samples are prepared before the training using some other MCMC algorithms, while for the reverse KL (which is our case), the samples are drawn from the model itself.} while also proposed several mitigation strategies. We have tested two methods from Ref.~\cite{Hackett:2021idh} that seem to provide better results:
\begin{description}[style=unboxed,leftmargin=0.5cm]
    \item[$\bullet$~Adiabatic retraining:] The idea of adiabatic retraining is derived from transfer learning, wherein a model trained targeting some easier-to-approximate set of parameters is used to initialize the training for harder ones. Adiabatic retraining generalizes this idea to vary the parameters epoch-by-epoch over some trajectory through parameter space over the course of training. As long as the variation is sufficiently slow, the model being trained will remain a good approximation of the target distribution throughout training. The way in Ref.~\cite{Hackett:2021idh} applies this method is to start the training from modeling a unimodal distribution and then adiabatically transform it into a multimodal one, which requires certain knowledge of the target distribution before training.
    \item[$\bullet$~Flow-distance regularization:] The flow-distance regularizer is a term added to the loss function to penalize flow functions $f$ that transform prior samples $z$ to significantly different output samples $\phi$, as given by
    \begin{equation}
        L(\tilde{p},p)_{\rm fd} = L(\tilde{p},p) + k\,[1-g(t)]\,\frac{1}{D}\sum_{i=1}^D\Vert \phi_i-z_i\Vert_2 ~,
    \end{equation}
    where the $L^2$ or $\Vert \Vert_2$ norm implies a sum over all degrees of freedom on the lattice, $k$ is the coupling strength of this constraint, and $g(t)\in[0,1]$ is a scheduler function of training time $t$ that gradually turns off the regularization. To some extent, this method is applicable to any theory without a priori knowledge of the mode structure and relies only on the broadness of the prior.
\end{description}
In practice, one can combine these two methods as they are both $t$-dependent. Ref.~\cite{Hackett:2021idh} has shown that this method can avoid the problem of undersampling a minimum for a (slightly-broken) $\mathbb{Z}_2$-symmetric $\phi^4$ theory compared to the other flow-based methods that intrinsically relies on the $\mathbb{Z}_2$ symmetry. 

In the context of cosmological phase transition, however, the relative weights among the stable and metastable vacua as well as the critical-bubble configuration are usually more than several orders of magnitude, making the problem almost unsolvable using even the revised flow-based multimodal modeling method mentioned above. There are at least two reasons for this: 1) The large hierarchy among the probability weights makes challenge $\mathcal{B}$ or the mode-collapse problem much more serious and cannot be dealt with easily. 2) More fundamentally, even if we have a perfectly trained flow model, it still cannot sample from the minimum of the probability distribution efficiently, as described by challenge $\mathcal{C}$ or the ``rare-event sampling'' problem.

To address these challenges, Ref.~\cite{Moore:2000jw} uses the method of MMCMC (as briefly mentioned at the end of Sec.~\ref{subsec:nucleation:lattice}), which is in essence a reweighted sampling method based on a modification to the action
\begin{equation}
    S_{\rm E}[\phi] \to S_{\rm E}[\phi] + W\big(\mathcal{O}(\phi)\big) ~,
\end{equation}
that aims to uniformly sample across the measurable $\mathcal{O}(\phi)$ space by choosing $W\big(\mathcal{O}(\phi)\big)\approx-\log P\big(\mathcal{O}(\phi)\big)$. Here, $P\big(\mathcal{O}(\phi)\big)$ is the true probability distribution of $\mathcal{O}(\phi)$. After that, one can retrieve the correct weights by performing a post-reweighting. However, one cannot in principle know $W\big(\mathcal{O}(\phi)\big)$ a priori. Ref.~\cite{Moore:2000jw} resolves this issue by performing two rounds of sampling: in the first round, a piecewise construction of $W\big(\mathcal{O}(\phi)\big)$ is performed on the fly during an evolving MMCMC sampling process; after the first round, one then uses the constructed $W\big(\mathcal{O}(\phi)\big)$ to perform a full MMCMC sampling. More recently, Ref.~\cite{gao2023rare} proposes the ``normalizing flow assisted importance sampling (NOFIS)'' algorithm to estimate the rare event probability. The basic idea is similar to that of MMCMC, which aims to train a NF model based on a reweighted target distribution. Given that the region of rare events to be modeled is known a priori (but it is either hard or time-consuming to compute the associated probability directly), one can divide the region into a series of nested sub-regions with strictly decreasing probabilities. 
First, one should focus on modeling the smallest sub-region by ``suppressing" the probabilities outside. Based on this initial model, one can then proceed to model the next-level sub-region. This process is repeated until the entire rare-event region is properly modeled.

\subsection{Partitioning flow-based MCMC}\label{subsec:method:slice}

Inspired by the ideas of MMCMC and NOFIS, we propose the method of ``partitioning flow-based (PF) MCMC'' (PFMCMC) to address the critical bubble sampling problem. Similar to MMCMC and NOFIS, we train our flow models based on reweighted actions; however, instead of trying to model the target distribution globally, we ``slice'' the space of $\mathcal{O}(\phi)$ into several partitions and then ``glue'' them together to obtain the ``overall'' $P\big(\mathcal{O}(\phi)\big)$. Furthermore, since we cannot identify the rare-event region a priori (and thus we cannot directly adopt the NOFIS method), our slicing is done in a sequential rather than in a nested fashion for the purpose of searching for the minimum.

There are at least three advantages of this approach: 1) If chosen wisely, all of the partitions will be either unimodal (around the two vacua) or monotonic except for the one containing the critical-bubble configuration, which, however, should have a small enough hierarchy that can be properly managed. Therefore, within each partition of the order-parameter space, flow-based MCMC can easily outperform traditional MCMC as has been demonstrated in many previous works. 2) Compared to modeling a multimodal distribution using one single flow model directly, modeling several unimodal and/or monotonic distributions (plus a small-hierarchy U-shaped distribution) can in general be more efficient and stable. 3) When it comes to bubble nucleation rate calculation, we only need the information between $\mathcal{O}_F$ and $\mathcal{O}_C$, and thus we can ignore much of the order-parameter space from around $\mathcal{O}_C$ to $\mathcal{O}_T$ using this method, even though there is no obstacle to construct the remaining space. 
The flip side of our method is that we will need to train multiple models for one single benchmark. On the other hand, since most of the partitions will either be monotonic or unimodal, the training of each model will be easier to be managed.~\footnote{We also remark that one can try to construct a mixture model based on the sub-models of these partitions as mentioned in Ref.~\cite{Hackett:2021idh}, but that is not necessary for the calculation of bubble nucleation rates.}

We now explicitly explain the methodology. For each model benchmark, we choose a wide range of the simulation region in terms of the measurable  $[\mathcal{O}_L,\mathcal{O}_R]$ to include both $\mathcal{O}_F$ and $\mathcal{O}_C$, or $\mathcal{O}_F,\mathcal{O}_C\in[\mathcal{O}_L,\mathcal{O}_R]$.
We partition the simulation region $[\mathcal{O}_L,\mathcal{O}_R]$ into 
several mildly hierarchical overlapping partitions with each individual partition defined by $[\mathcal{O}_i,\mathcal{O}_j]$. Targeting one individual partition, we train a flow model according to the reweighted action
\begin{equation}
    S_{\rm E}^{ij}[\phi] = \begin{cases}
        S_{\rm E}[\phi] + W_{\rm L}^i\times\left(e^{B_{\rm L}^i(\mathcal{O}_i - \mathcal{O})} - 1\right) \,,\qquad \mathcal{O} \leq \mathcal{O}_i \vspace{2mm} \\ 
         S_{\rm E}[\phi] \,, \qquad \qquad \qquad \qquad \qquad \qquad \quad \mathcal{O}_i<\mathcal{O}(\phi)<\mathcal{O}_j \vspace{2mm} \\ 
        S_{\rm E}[\phi] + W_{\rm R}^j\times \left(e^{B_{\rm R}^j(\mathcal{O} - \mathcal{O}_j)} - 1\right) \,,\qquad  \mathcal{O}_j \leq \mathcal{O}
    \end{cases} ~,
\end{equation}
where $W_{\rm L}^i,W_{\rm R}^j,B_{\rm L}^i,B_{\rm R}^j>0$ are manually chosen parameters to ``suppress'' the irrelevant order-parameter space, thus effectively making a ``partition''. We note that the choices of the $W$ and $B$ parameters are not sensitive to the lattice setup, and are acceptable as long as the sideband regions can be suppressed effectively while the numerical stability of the computation is realized. After sampling from each partition, we glue the partitions together following this procedure: 
\begin{enumerate}
    \item Starting from the partition containing the local minimum, we simply take the histogram values of the partition in focus within the non-overlapping regions.
    \item When two partitions overlap, the endpoint of one of the partitions in that region must be a (local) maximum of the original partition. We denote the probability distribution of this first partition (partition 1) as $P_1(\langle\phi\rangle)$, while the other (partition 2) as $P_2(\langle\phi\rangle)$. We scale the entire partition 2 so that the endpoint of partition 1 matches to the value at the same point of partition 2.
    \item For the remaining points in the overlapping region, we take a simple weighted average of the histogram values from the two partitions $[w_1P_1(\langle\phi\rangle)+w_2P_2(\langle\phi\rangle)]/(w_1+w_2),w_1\geq w_2$. In our study, this is done by human intervention, while a more sophisticated gluing algorithm is certainly possible.
\end{enumerate}
The idea behind steps 2. and 3. is that it is more likely for a flow model to undersample or mismodel a minimum than a maximum, which is a compelling reason for making overlapping partitions than mutually exclusive ones. 

We note that the idea of ``cutting'' has been adopted in a different way in Ref.~\cite{Shen:2022xcm} to calculate the vacuum decay rate of a quantum mechanical system using the ``wait-and-see approach''. In their work, they measure the similarity of the sampled configurations to the true vacuum with a ``modified'' action integral that only considers the classically allowed region in the solution space. By imposing a hard cut on this integral, they are able to discard the phase space near the true vacuum. In contrast, we apply the technique in the context of Langer's formalism for relativistic quantum fields. Also, instead of making the cuts based on the action, our algorithm partitions the order parameter space and can thus directly identify the different regions of the vacuum structure.

\subsubsection{A toy $2$D $\phi^4$ theory}\label{subsubsec:2Dphi4}

\begin{figure}[ht!]
    \centering
    \includegraphics[width=0.6\textwidth]{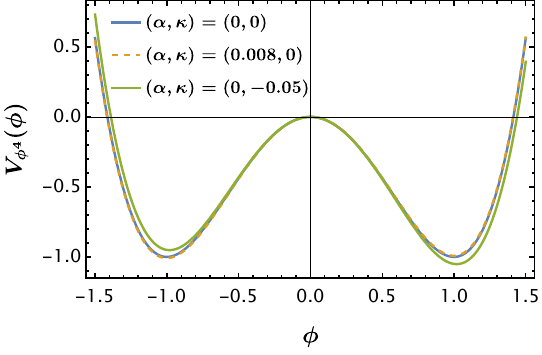}
    \caption{The potential profiles of the three model benchmarks of the 2D $\phi^4$ theory [see Eq.~\eqref{eq:phi-4}]. Here, all benchmarks have the common choice of $(m^2,\lambda)=(-2,1)$, with $(\alpha,\kappa)=$ (I) $(0,0)$, (II) $(0.008,0)$, and (III) $(0,-0.05)$, respectively.}
    \label{fig:V-f4} 
\end{figure}

As an example, we use a toy $2$D (only two spacial dimensions) $\phi^4$ theory on a lattice of size $L\times L$ to illustrate our PFMCMC algorithm. The action of this model is described by
\begin{equation}\label{eq:phi-4}
    S_{\phi^4}[\phi] = \int d^2x\left[\frac{1}{2}(\partial^\mu\phi)^2 + V_{\phi^4}(\phi)\right] ,\quad  V_{\phi^4}(\phi) = \alpha \, \phi + m^2\, \phi^2 + \kappa\, \phi^3 + \lambda\, \phi^4 ~,
\end{equation}
where the discretized Laplacian operator is given to the leading order by
\begin{equation}\label{eq:laplacian}
    (\nabla_{\rm L}^2)_{x}^{(1)} = \sum_{i=a\hat{x},a\hat{y}} \frac{1}{a^2}\left(\phi_{x+i}+\phi_{x-i}-2\phi_x\right) ~.
\end{equation}
The measurable $\mathcal{O}$ is chosen to be the spacial mean field value or 
\beqa
\mathcal{O}(\phi) =\langle\phi\rangle\equiv \frac{1}{L^2}\,\int d^2x \, \phi(x) ~.
\eeqa
We choose three model benchmarks with the common choice of $(m^2,\lambda)=(-2,1)$, and with (I) $(\alpha,\kappa)=(0,0)$, (II) $(\alpha,\kappa)=(0.008,0)$, and (III) $(\alpha,\kappa)=(0,-0.05)$, respectively. Their potential profiles are shown in Fig.~\ref{fig:V-f4}.

For the flow model, we use real-valued non-volume preserving (real NVP) flows~\cite{dinh2017density} constructed from non-linear squared couplings (NLSq)~\cite{ziegler2019latent}
\begin{equation}
    g(x) = a + b\,x + \frac{c}{1+(d\,x+g)^2} ~,
\end{equation}
where $a,b,c,d,g$ are model parameters to be optimized. Following Ref.~\cite{Albergo:2019eim}, each coupling layer checkerboards the sites and updates one parity (white or black) of sites conditioned on the frozen values of the other parity, with the parities switched between layers (see Fig.~\ref{fig:NF-cartoon}). We use a convolutional neural network (CNN) of 4 hidden layers with 12 nodes and a kernel size of 5 to parametrize the transformations in each coupling layer. All hidden layers have a leaky ReLU activation of negative slope 0.01, \ie
\begin{equation}
    {\rm Leaky~ReLU}_{0.01}(x) = \begin{cases}
        x ,~\quad \quad \quad  x > 0 \\
        0.01\,x ,~\quad {\rm otherwise}
    \end{cases} ~,
\end{equation}
except for the last one which has a $\tanh$ activation. The entire model consists of 40 such coupling layers, resulting in roughly 500K trainable hyperparameters. We choose the lattice size $L=10$, lattice spacing $a=1$, and batch size $N=1000$. We split the $\langle\phi\rangle$ space into three partitions: $[-1.5, 0]$, $[-0.5, 0.5]$, and $[0, 1.5]$, with a common choice of $W_{\rm L}^i=W_{\rm R}^i=1000$ and $B_{\rm L}^i=B_{\rm R}^j=1$. We train the models based on the loss function given in Eq.~\eqref{eq:reverse-KL}.

\begin{figure}[ht!]
    \centering
    \includegraphics[width=0.47\textwidth]{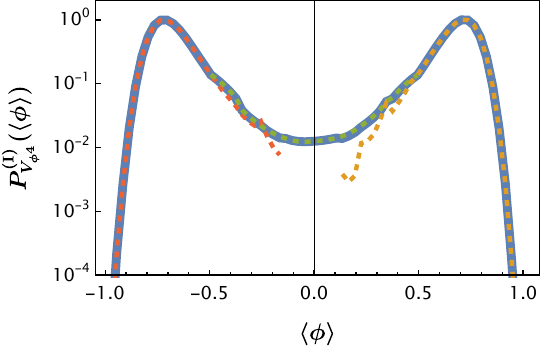} \hspace{3mm}
    \includegraphics[width=0.47\textwidth]{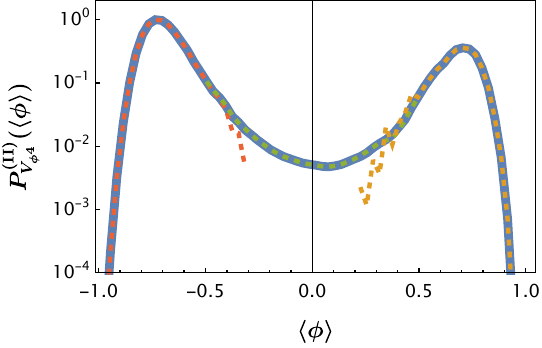}
    \caption{{\it Left:} The simulated probability distribution $P(\langle \phi \rangle)$ for benchmark (I) of the $V_{\phi^4}$ model is shown in the blue and solid line and is normalized by the value at the (global) maximum. The red, green and yellow dashed lines are the distributions of the individual partitions. {\it Right:} The same as the left one, but for benchmark (II).
    }
    \label{fig:b12}
\end{figure}

\begin{figure}[th!]
    \centering
    \includegraphics[width=0.47\textwidth]{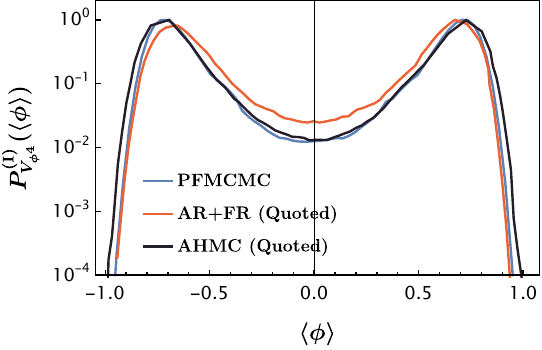} \hspace{3mm}
    \includegraphics[width=0.47\textwidth]{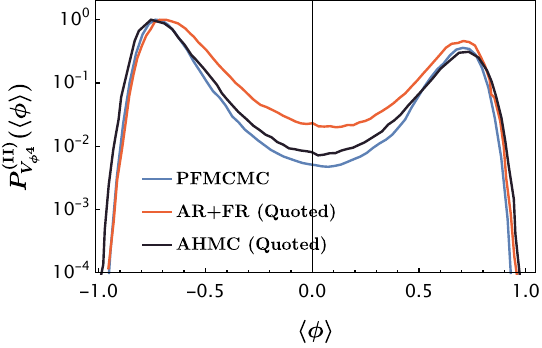}
    \caption{{\it Left:} The probability distributions of benchmark (I) obtained through the PFMCMC, AR+FR, and AHMC methods. {\it Right:} The same as the left one, but for benchmark (II).  The AR+FR and AHMC distributions are directly quoted from Ref.~\cite{Hackett:2021idh}.
    }
    \label{fig:comp}
\end{figure}

Using the PFMCMC method described above, we first show the probability distributions of $\langle\phi\rangle$ [normalized by the value at the (global) maximum] for benchmarks (I) and (II) of the 2D $\phi^4$ model in Fig.~\ref{fig:b12}. 
For this toy model, we compare the distributions of benchmarks (I) and (II) obtained through our method to those through the existing two methods reported in Ref.~\cite{Hackett:2021idh} in Fig.~\ref{fig:comp}: the adiabatic retraining + flow-distance regularization (AR+FR) method and the augmented hybrid MC (AHMC) method, the latter of which can be thought of (not guaranteed) as the ``truth distributions''.~\footnote{
One can view the AHMC method as directly sampling from the target distribution, whereas the other two methods (PFMCMC and AR+FR) might exhibit bias toward the proposal distributions learned by the flow models. Therefore, if AHMC is executed perfectly, it should generate the ``truth distributions," yet in principle, there is no analytical means to confirm this.
} 
We can see that for both benchmarks, the PFMCMC distributions are very well aligned with the AHMC distributions except for the local minimum of benchmark (II), which is slightly underestimated.  This latter issue can in principle be resolved by making finer partitions. On the contrary, the AR+FR method consistently overestimates the local minima. These comparisons demonstrate the success of our PFMCMC algorithm to obtain the right probability distributions and the reliability of the PFMCMC method when constructing multimodal distributions.

\begin{figure}[ht!]
    \centering
    \includegraphics[width=0.47\textwidth]{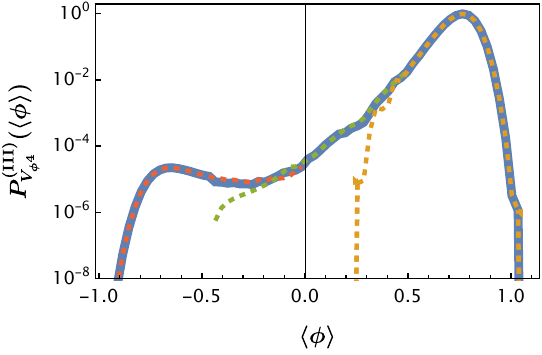}
    \caption{The same as Fig.~\ref{fig:b12} but for benchmark (III). 
    }
    \label{fig:b3}
\end{figure}

We further show the probability distribution of benchmark (III) obtained through the PFMCMC method in Fig.~\ref{fig:b3}. For this benchmark point, the probability of the metastable vacuum is suppressed by $\sim\mathcal{O}(10^{-5})$ compared to the stable vacuum. Our PFMCMC method can still capture the bimodal structure, which is difficult to obtain using the straightforward flow-based sampling methods.

\subsubsection{A toy $2$D $\phi^6$ theory}\label{subsubsec:2Dphi6}
\begin{figure}[ht!]
    \centering
    \includegraphics[width=0.47\textwidth]{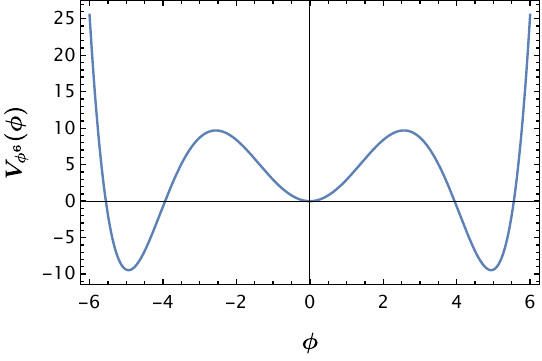}
    \caption{The potential profile of the 2D $\mathbb{Z}_2$-symmetric $\phi^6$ theory [see Eq.~\eqref{eq:phi-6}] benchmark with $(m^2,\lambda,\eta)=(6.50,-7.53,4.87)$.}
    \label{fig:V-f6}
\end{figure}

As a demonstration of how the PFMCMC method can be used to model a largely hierarchical distribution and as a prelude to our study of the finite-temperature (2+1)D scalar field theory in Sec.~\ref{sec:sampling}, we study one benchmark of the 2D $\mathbb{Z}_2$-symmetric $\phi^6$ theory described by
\begin{equation}\label{eq:phi-6}
    S_{\phi^6}[\phi] = \int d^2x\left[\frac{1}{2}(\partial^\mu\phi)^2 + V_{\phi^6}(\phi)\right] ,\quad \quad V_{\phi^6}(\phi) = \frac{m^2}{2}\phi^2 + \frac{\lambda}{4!} \phi^4 + \frac{\eta}{6!} \phi^6 ~,
\end{equation}
with $(m^2,\lambda,\eta)=(6.50,-7.53,4.87)$. The potential profile is shown in Fig.~\ref{fig:V-f6}. We still choose the same measurable $\mathcal{O}(\phi)=\langle\phi\rangle$, but normalize the probability distribution by the value at $\langle\phi\rangle=0$, which is the metastable vacuum. Moreover, while this benchmark has a two-fold stable vacuum degeneracy, we only focus on transition to the positive stable vacuum. The same order parameter, normalization scheme, and choice of the stable vacuum are also used in the later study of the finite-temperature theory (see Sec.~\ref{sec:sampling}).
For this particular benchmark, we choose a slightly smaller lattice size $L=8$ with the same lattice spacing $a=1$. Because this benchmark has a probability hierarchy of $\sim\mathcal{O}(10^{-50})$ between the metastable vacuum and the critical-bubble configuration, we deploy roughly 20 partitions to apply the PFMCMC method. For each partition, we adopt the same model and train upon the same loss function as before and choose $W_{\rm L}^i=W_{\rm R}^i=1000$, $B_{\rm L}^i=6$, and $B_{\rm R}^j=4$. The different choices for $B_{\rm L,R}$ is because we only focus on scanning the region between the metastable vacuum and the critical configuration and thus do not expect as much effect from the stable vacuum.

\begin{figure}[ht!]
    \centering
    \includegraphics[width=0.47\textwidth]{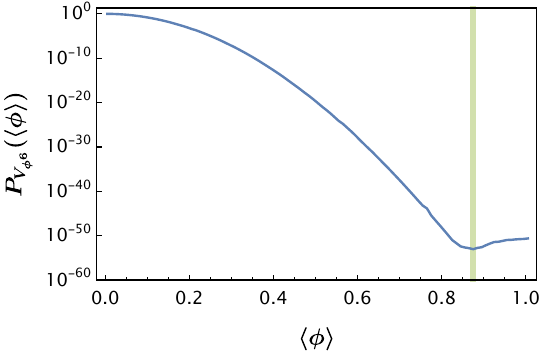} \hspace{3mm}
    \includegraphics[width=0.47\textwidth]{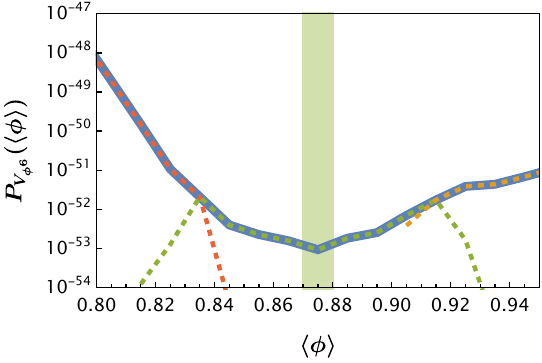}
    \caption{{\it Left:} The overall view of the $P_{V_{\phi^6}}(\langle\phi\rangle)$ distribution [normalized by $P_{V_{\phi^6}}(0)$] for the $V_{\phi^6}$ model benchmark. {\it Right:} The zoom-in view of the left panel around the critical separatrix. The green band given by $\langle\phi\rangle\in[0.870,0.880]$ stands for the critical separatrix. The dashed lines in the right panel stand for the individual partition distributions while the blue and solid line for the combined distribution. The sharp drops at the endpoints of the green and red dashed curves are the results of partitioning, whose values are not used in the analysis.}
    \label{fig:mean-f6}
\end{figure}

We show the simulated final distribution $P_{V_{\phi^6}}(\langle\phi\rangle)$ (left panel) and the zoom-in view around the critical separatrix (right panel) in Fig.~\ref{fig:mean-f6}. In the right panel, we show both the individual partition and combined distributions. We note that the sharp drops at the end points of the green and red dashed curves are the results of partitioning, whose values are not used in the analysis. For this benchmark, we choose the critical separatrix to be $\langle\phi\rangle\in[0.870,0.880]$ in accordance with the histogram bin width, with the local minimum sitting at $\langle\phi\rangle=0.875$ and $\epsilon=0.01$. The statistical factor of this particular benchmark is given by 
\begin{equation*}
    F_{\rm S} = \dfrac{1}{2}\times \dfrac{P_{V_{\phi^6}}\big(\langle\phi\rangle\in[0.870,0.880]\big)}{0.01\times P_{V_{\phi^6}}\big(\langle\phi\rangle\in[0,0.870]\big)} = 8.60\times10^{-53}~,
\end{equation*}
where the factor of $1/2$ accounts for the negative side of the metastable vacuum for this $\mathbb{Z}_2$-symmetric model. We remark that the large hierarchy in this benchmark is unlikely to be captured using the straightforward flow-based sampling methods introduced before.

\section{(2+1)D $\mathbb{Z}_2$-symmetric real scalar theory}\label{sec:sampling}

We apply the PFMCMC method to study the finite-temperature (2+1)D $\mathbb{Z}_2$-symmetric real scalar theory. We first summarize the properties of the theory and choose one benchmark with a manageable probability hierarchy. Then, we calculate the statistical factor of the benchmark using the PFMCMC method with a few different lattice setups. To demonstrate the nucleation rate calculation procedure, we choose the setups with unit lattice spacing $a=1$ and calculate the corresponding dynamical factors.
After that, we ``compare'' the calculated nucleation rates of the benchmark to the one obtained using \texttt{CosmoTransitions}~\cite{Wainwright:2011kj} and \texttt{BubbleDet}~\cite{Ekstedt:2023sqc}, packages based on the bounce solution action using the saddle-point approximation, in a way that takes into consideration the finite-spacing effect.

\subsection{Finite-temperature (2+1)D $\mathbb{Z}_2$-symmetric theory}\label{subsec:sampling:thermal}

The finite-temperature Euclidean action of the system is defined in a 2D space and is given by
\begin{equation}\label{eq:2+1:action}
    S_{\rm E}[\phi] = \frac{1}{T}\int d^{2}x\left[\frac{1}{2}\partial_\mu \phi \partial^\mu \phi + V_{\rm eff}(\phi)\right] \equiv \frac{S[\phi]}{T} ~.
\end{equation}
Here, the finite-temperature effective potential $V_{\rm eff}(\phi)$ is calculated at one-loop level (see Appendix~\ref{sec:app-finite-temp} for detailed calculations) and is
\beqa
V_{\rm eff}(\phi) = V_0(\phi) + V_1(\phi,T) + V_\beta(\phi,T) ~,
\eeqa
with
\begin{equation}
    V_0(\phi) = \frac{m^2}{2}\phi^2 + \frac{\lambda}{4!}\phi^4 + \frac{\eta}{6!}\phi^6 ~, \qquad 
    V_1(\phi,T)  = -\frac{1}{12\pi}m_{\rm eff}^3(\phi,T) ~,
\end{equation}
\begin{equation}
    V_\beta(\phi,T) = -\frac{T^3}{2\pi}\left[\frac{m_{\rm eff}}{T}\,{\rm Li}_2\left(e^{-\frac{m_{\rm eff}}{T}}\right) + {\rm Li}_3\left(e^{-\frac{m_{\rm eff}}{T}}\right)\right] ~,
\end{equation}
with ${\rm Li}_{n}$ denoting the polylogarithmic functions. Here $m_{\rm eff}$ stands for the thermal effective mass and is given by
\begin{equation}
    m_{\rm eff}^2(\phi,T) = m_T^2 + \frac{\lambda}{2}\,\phi^2 + \frac{\eta}{4!}\,\phi^4 + \frac{\eta}{4}\,\phi^2\,\tau_1(T,m_T) ~,
\end{equation}
where $m_T$ is the IR-cutoff parameter and
\begin{equation}
    \tau_1(T,m_T) = -\frac{T}{2\pi}\ln\left(2\sinh\frac{m_T}{2\,T}\right) ~.
\end{equation}
Though the non-analytic structure of the potential can induce imaginary components, the benchmark that we choose for the remaining study is strictly real, as we will demonstrate in the next section. The details of the theory can be found in Appendix~\ref{sec:app-finite-temp}. While this theory has a two-fold degeneracy in the stable vacua, we only focus on the transition to the positive vacuum in the remaining study.

\subsection{Benchmark test: bubble simulation}\label{subsec:sampling:finite}

\begin{figure}[ht!]
    \centering
    \includegraphics[width=0.47\textwidth]{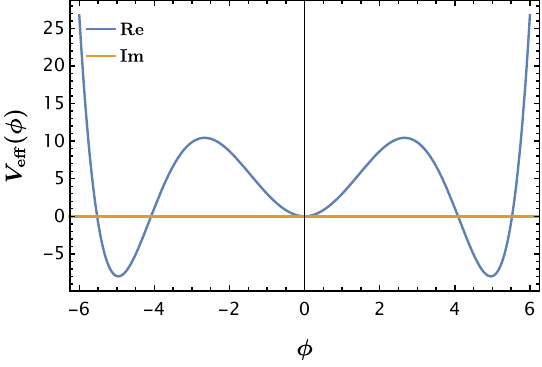} \hspace{3mm}
    \includegraphics[width=0.47\textwidth]{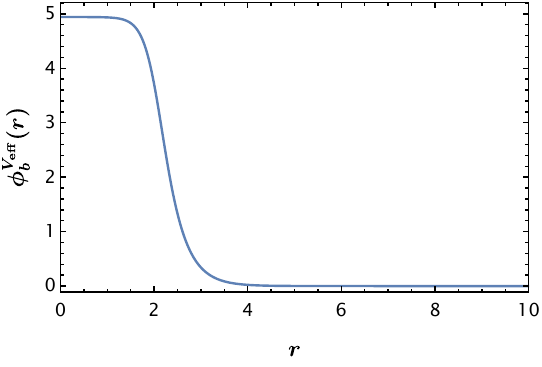}
    \caption{{\it Left:} the potential profile for the benchmark point of the finite-temperature effective potential of the (2+1)D $\mathbb{Z}_2$ real scalar theory with $(m^2,\lambda,\eta,T)=(1,-5,10,7.2)$, which corresponds to $m_T^2\approx1.296$. Note that the imaginary component of the potential is zero. {\it Right:} the corresponding bounce solution for this benchmark potential. The bounce action is $S_{b, {\rm eff}}/T \approx 16.60$.
    }
    \label{fig:bounce:eff}
\end{figure}

For the benchmark test, we choose $(m^2,\lambda,\eta,T)=(1,-5,10,7.2)$,~\footnote{We keep the unit of energy to be 1, \ie, $m=1$.} which corresponds to $m_T^2\approx1.296$. The potential profile and the corresponding bounce solution are shown in Fig.~\ref{fig:bounce:eff}. From the bounce solution, one can see that the spherical bubble radius $r_{b,{\rm eff}}\approx 2.5-3$. The bounce action of this benchmark is $S_{b,{\rm eff}}/T\approx 16.60$. We remark that for this benchmark, the potential is strictly real, as can be seen from the imaginary component plotted in the figure. Also, this benchmark is chosen to have a smaller probability hierarchy in the measurable space for the sake of simulation efficiency (there should be no obstacle to simulate a benchmark with a more hierarchical probability distribution, as we demonstrated in Sec.~\ref{subsec:method:slice}). As a proof of concept of the PFMCMC method, our discretization scheme simply takes the first-order discretized Laplacian given in Eq.~\eqref{eq:laplacian} and uses the continuum effective potential for the lattice action. For a rigorous study of the lattice perturbation theory of scalar fields, one can refer to Ref.~\cite{Gould:2021dzl}.

For the PFMCMC simulation, we use the same model and loss function as the ones described in Sec.~\ref{subsec:method:slice}. In principle, a proper choice of $a$ should be much smaller than the correlation length $\xi$ of the system, \ie, $a \ll \xi\sim 1/m_\phi$, $m_\phi$ being the mass of the scalar field~\cite{Gould:2021dzl}. Furthermore, one should extract the continuum limit result by sampling a few different $a$'s and taking $a\to0$ based on a function fit (such as $b_0+b_1a^3$ used in Ref.~\cite{Gould:2024chm}). In a system with a finite lattice spacing, the high-momentum modes are cut off, leading to a hard upper bound on the bubble wall tension scale. In the thin-wall limit, the critical-bubble radius scales roughly as $\sigma/\Delta V$, $\sigma$ being the one-dimensional wall tension and $\Delta V$ the two-dimensional vacuum pressure difference, and thus one direct outcome of a finer lattice spacing is the increase of the bubble radius, which will become much milder when the previous condition $a\ll \xi$ is satisfied. This requirement will in general lead to a high training cost of the flow models, which we will discuss in Sec.~\ref{sec:conclusions}. Since this work is a proof of concept of the PFMCMC method, we will not restrict ourselves to the mentioned condition, but will only demonstrate the finite spacing effect on the simulation results without performing any function fits or extrapolations. 

\begin{figure}[th!]
    \centering
    \includegraphics[width=0.48\textwidth]{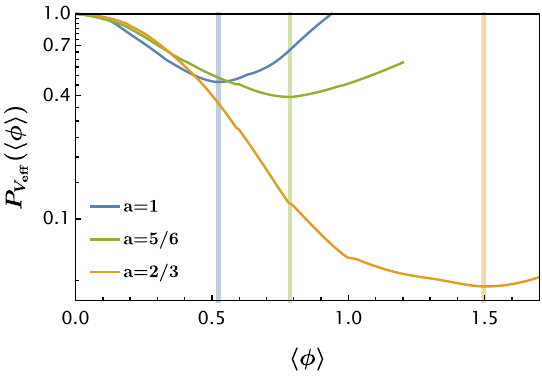}
    \caption{
    The $P_{V_{\rm eff}}\big(\langle\phi\rangle\big)$ distributions for $L=10$ and $a=1,\frac{5}{6}, \frac{2}{3}$, respectively. The critical separatrices are shown in terms of the colored bands with the numerical values given in Eq.~\eqref{eq:critical:sizes:a}.
    }
    \label{fig:prob:FT:a}
\end{figure}

\begin{figure}[th!]
     \centering
     \includegraphics[width=0.80\textwidth]{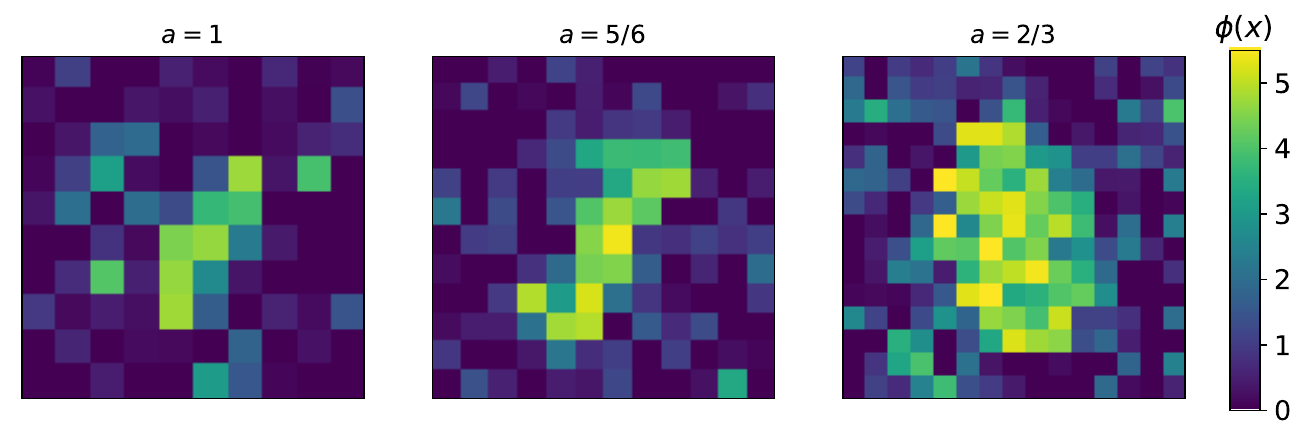}
     \caption{Example critical-bubble configurations properly contained in the lattice of $L=10$ and ({\it left}) $a = 1$, ({\it middle}) $a = 5/6$, and ({\it right}) $a=2/3$, respectively. Their $\langle\phi\rangle$ values fall within the critical separatrices defined in Eq.~\eqref{eq:critical:sizes:a}.
     }
     \label{fig:bubbles:a}
 \end{figure}

For this purpose, we fix the lattice size $L=10$ and choose a few different lattice spacings $a=1, \frac{5}{6}, \frac{2}{3}$. We deploy $1,1,2$ partitions for the three lattice spacings within the ranges of $\langle\phi\rangle\in[0,1],[0,1.2],[0,1.0]\cup[0.7,1.7]$, respectively, with a common choice of $W_{\rm L}^i=W_{\rm R}^i=1000$ and $B_{\rm L}^i=B_{\rm R}^i=2$ for all partitions. The calculated probability distributions with respect to $\langle\phi\rangle$ [$P_{V_{\rm eff}}\big(\langle\phi\rangle\big)$] are shown in Fig.~\ref{fig:prob:FT:a}. The critical separatrices are given by 
\beqa\label{eq:critical:sizes:a}
\langle\phi\rangle_{a=1}\in[0.518,0.528] \,,~\quad
\langle\phi\rangle_{a=\frac{5}{6}}\in[0.781,0.791] \,,~\quad
\langle\phi\rangle_{a=\frac{2}{3}}\in[1.491,1.501] ~.
\eeqa
It can be seen that the position of the critical separatrix shifts to the right as $a$ decreases, which implies the increase of the critical-bubble size. Furthermore, the probability hierarchy also increases along with the decrease in $a$, which is another result of including higher-momentum modes as they can contribute to larger free energy differences. We also show one critical-bubble configuration for each of the three setups in Fig.~\ref{fig:bubbles:a}, from which one can see the increase in the bubble size as $a$ decreases.

On the other hand, to determine a proper lattice size, we follow the argument given in Ref.~\cite{Moore:2000jw} to derive the condition. First, the lattice size $L$ should be larger than the bubble size or $L > 2r_b$ to contain the whole bubble. Here $r_b$ stands for the radius of a generic simulated bubble and is measured through real simulations, which could differ from the one obtained from the bounce solution profile, $r_{b,{\rm eff}}$ (see the right panel of Fig.~\ref{fig:bounce:eff}). Furthermore, to avoid the dominance of unphysical periodic ``bubbles'' (bubbles which will minimize their free energies by connecting through the unphysical periodic boundaries), one needs a more stringent condition: $L\gtrsim \pi r_b$. Nevertheless, for a too large $L$, the probability of obtaining the interesting bubble configurations will be diluted by noises outside the bubble region.~\footnote{See Ref.~\cite{Gould:2024chm} for a better choice of the measurable to address this issue.} It has been argued and shown in Refs.~\cite{Moore:2000jw,Gould:2022ran} that as long as the lattice size satisfies the previous condition, the nucleation rate should only have exponentially small volume-dependence, which can be shown through a fit based on the function $\Gamma_\infty+c_0\exp(-c_1 L)$ with $c_{0, 1}>0$, and taking $L\to\infty$ will give the infinite-volume result. Moreover, because of the fixed-size nature of the critical bubbles, another signature of a properly chosen lattice size is the mildly volume-dependent distribution of the volume-integrated measurables, such as $L^2\langle\phi\rangle$. For demonstration purpose, we choose three different lattice sizes $L=9,10,11$ that are not significantly larger than $\pi r_b$ with unit lattice spacing $a=1$, for which we find that $r_b\lesssim 3$ for the majority of bubbles. We will explicitly show the mild volume-dependence of the $L^2\langle\phi\rangle$ distribution, but will not perform any function fits or extrapolations.

\begin{figure}[ht!]
    \centering
    \includegraphics[width=0.48\textwidth]{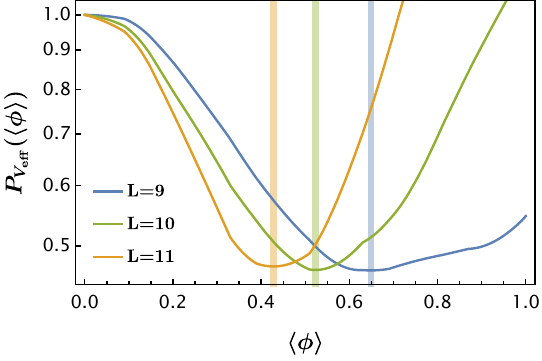}
    \includegraphics[width=0.48\textwidth]{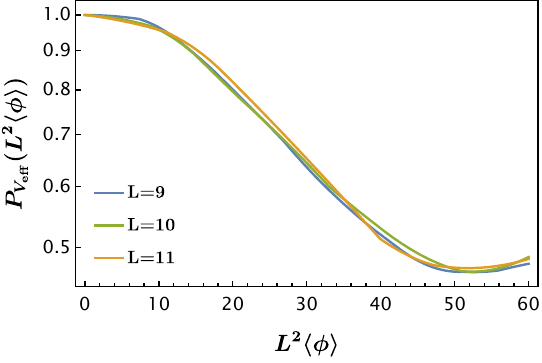}
    \caption{{\it Left:} The $P_{V_{\rm eff}}(\langle\phi\rangle)$ distributions for $a=1$ and $L=9,10,11$, respectively. {\it Right:} The $P_{V_{\rm eff}}(L^2\langle\phi\rangle)$ distributions for the same lattice setups. The critical separatrices are shown in terms of the colored bands in the left plot with the numerical values given in Eq.~\eqref{eq:critical:sizes}.
    The curves in the right panel are mildly volume-dependent because $L^2\langle\phi\rangle$ should have a fixed value for a fixed-size bubble.
    }
    \label{fig:prob:FT}
\end{figure}

\begin{figure}[ht!]
     \centering
     \includegraphics[width=0.55\textwidth]{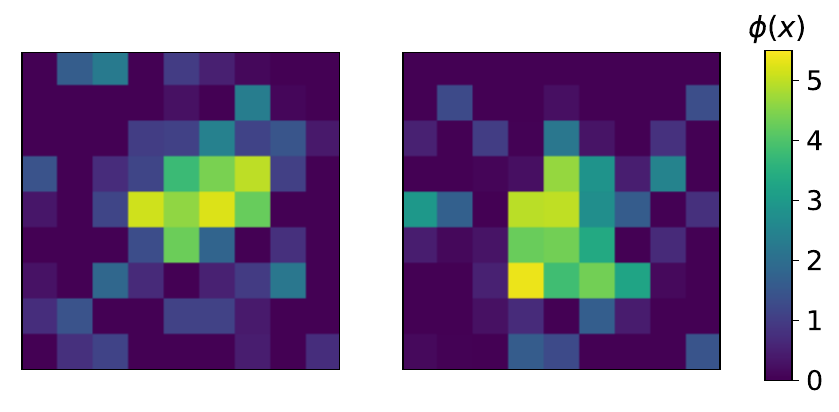}
     \caption{Example critical-bubble configurations properly contained in the lattice of $L=9$ and $a = 1$. Their $\langle\phi\rangle$ values fall within the critical separatrix defined in Eq.~\eqref{eq:critical:sizes}.
     }
     \label{fig:bubbles}
 \end{figure}

For each lattice size, we deploy one partition of $\langle\phi\rangle\in[0,1]$ with the common choice of $W_{\rm L}^i=W_{\rm R}^i=1000$ and $B_{\rm L}^i=B_{\rm R}^i=2$. First note that $\langle\phi\rangle$ scales as $1/L^2$, so the partition $[0,1]$ can sufficiently cover the interesting region from $\mathcal{O}_F$ to $\mathcal{O}_C$. Secondly, this benchmark is not that hierarchic, so we do not need to choose many partitions. We show $P_{V_{\rm eff}}(\langle\phi\rangle)$ and the calculated probability distribution with respect to $L^2\langle\phi\rangle$ [$P_{V_{\rm eff}}(L^2\langle\phi\rangle)$] in Fig.~\ref{fig:prob:FT} for $L=9, 10, 11$, along with the chosen critical separatrices shown in the $P_{V_{\rm eff}}(\langle\phi\rangle)$ plot. The critical separatrices are given by 
\beqa\label{eq:critical:sizes}
\langle\phi\rangle_{L=9}\in[0.644,0.654] \,,~\quad
\langle\phi\rangle_{L=10}\in[0.518,0.528] \,,~\quad
\langle\phi\rangle_{L=11}\in[0.423,0.433] ~.
\eeqa
As pointed out in Refs.~\cite{Moore:2000jw,Gould:2022ran}, $P_{V_{\rm eff}}(L^2\langle\phi\rangle)$ is nearly identical between the false vacuum and the critical separatrix for different lattice sizes that satisfy the previously-mentioned minimal condition. The reason is that for a fixed-size bubble, $L^2\langle\phi\rangle$ is a fixed value and should exhibit mild volume dependence. For illustration purpose, we show two example critical-bubble configurations of the smallest lattice size $L=9$ in Fig.~\ref{fig:bubbles}, both of which are properly contained in the lattice. 

The statistical factors of different lattice sizes are calculated to be
 \begin{description}
     \item[$\bullet~L=9$] $$F_{\rm S} = \frac{1}{2}\times\frac{P_{V_{\rm eff},L=9}\big(\langle\phi\rangle\in[0.644,0.654]\big)}{0.01\times P_{V_{\rm eff},L=9}\big(\langle\phi\rangle\in[0,0.644]\big)} =0.502 ~.$$
     \item[$\bullet~L=10$] $$F_{\rm S} =\frac{1}{2}\times\frac{P_{V_{\rm eff},L=10}\big(\langle\phi\rangle\in[0.518,0.528]\big)}{0.01\times P_{V_{\rm eff},L=10}\big(\langle\phi\rangle\in[0,0.518]\big)} =0.622 ~.$$
     \item[$\bullet~L=11$] $$F_{\rm S} =\frac{1}{2}\times\frac{P_{V_{\rm eff},L=11}\big(\langle\phi\rangle\in[0.423,0.433]\big)}{0.01\times P_{V_{\rm eff},L=11}\big(\langle\phi\rangle\in[0,0.423]\big)} =0.762 ~.$$
 \end{description}
A simple power-law function $0.501\times (L/9)^{2.1}$ can provide a good fit to the three data points, whose $L^2$-dependence is mostly cancelled after being divided by $V=L^2$ [see Eq.~\eqref{eq:rate-full}]. The residual $L^2$-dependence is expected, and should (along with the $L^2$-dependence of the dynamical factors) be dealt with through the function fit mentioned previously.

\subsection{Dynamical factor and nucleation rate calculations}\label{subsec:sampling:rate}

To demonstrate the procedure of nucleation rate calculation, we finish the calculation of the dynamical factor using the setups with $a=1$ and $L=9,10,11$. Because the uncertainties of the statistical factors are much smaller than those of the dynamical factors, we will simply use their face values when calculating the nucleation rates. Since this paper mainly serves as a proof of principle to efficiently sample rare events and calculate the statistical factors, we will not perform the extrapolation to obtain the infinite-volume and continuum limits, which are necessary to calculate the nucleation rates of a realistic model.

The procedure to calculate the dynamical factor was described around Eq.~\eqref{eq:Langevin}. A proper timestep $\Delta t$ for a Langevin evolution should be shorter than the timescale to go from $\langle \phi\rangle_C$ to another configuration where $\log P(\langle\phi\rangle)$ differs by order 1. As long as this constraint is satisfied, the Brownian random walk nature of the evolution will guarantee that the quantity ${\rm S}\,\vert d\langle\phi\rangle/dt\vert_{\phi_C}$ is $\Delta t$-independent~\cite{Moore:2000jw}. To demonstrate this, we perform the Langevin evolution for the finite-temperature benchmark using the fourth-order Runge-Kutta method in real time with $\Delta t=0.010,0.015,0.020$.
On the other hand, we terminate the simulations when $\langle\phi\rangle < 0$ (enters the metastable phase) or $\langle\phi\rangle> \langle\phi\rangle_{\rm th}$ (enters the stable phase), with $\langle\phi\rangle_{\rm th}$ chosen such that the majority of the trajectories that have ``tunneled'' into the stable phase will not turn back and cross the critical separatrix again. To determine the proper $\langle\phi\rangle_{\rm th}$ for different lattice sizes, we try out a few different values and choose the smallest one for each size starting from which the dynamical factors obtained begin to converge. The explicit choices are $\langle\phi\rangle_{{\rm th}, (L=9)}=1.4$, $\langle\phi\rangle_{{\rm th}, (L=10)}=1.2$, $\langle\phi\rangle_{{\rm th}, (L=11)}=1.0$.

\begin{figure}[ht!]
    \centering
    \includegraphics[width=0.47\textwidth]{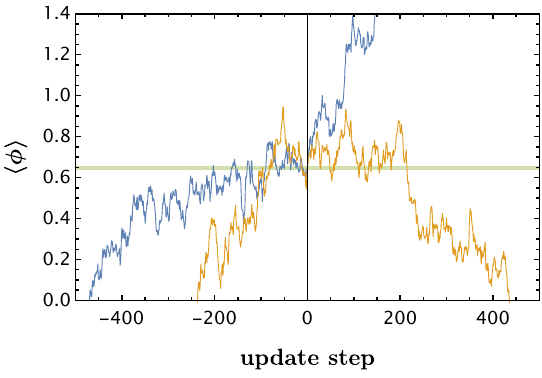}
    \caption{Two examples of Langevin evolution trajectories of the critical configurations with both forward and backward evolutions in time and ($L=9$, $a=1$, $\Delta t=0.01$). The blue trajectory has successfully tunnelled from the metastable to the stable vacuum, while the orange trajectory has failed. 
    The green band stands for the critical separatrix.}
    \label{fig:flux}
\end{figure}

We show two example trajectories of $L=9$, $a=1$ and $\Delta t=0.01$ in Fig.~\ref{fig:flux}, one of which successfully tunnels (even $N_{\rm crossings}$) while the other fails (odd $N_{\rm crossings}$). To calculate the dynamical factors, we generate roughly 2K critical configurations and collect 20 trajectories for each configuration for all $L$'s and $\Delta t$'s. The dynamical factors obtained are summarized in Table~\ref{tab:dynamical}. It can be seen that the dynamical factors of different $\Delta t$ do agree with each other within uncertainties.

\begin{table}[h!]
\renewcommand{\arraystretch}{1.5}
    \centering
    \begin{tabular}{|c|c|c|c|}
         \hline
             & $L=9$ & $L=10$ & $L=11$ \\
         \hline
         $\Delta t = 0.010$ & $(14.8\pm 1.22)\times 10^{-3}$ & $(11.1\pm 0.862)\times 10^{-3}$ & $(8.45\pm 0.751)\times 10^{-3}$ \\
         \hline
         $\Delta t = 0.015$ & $(14.5\pm 1.22)\times 10^{-3}$ & $(11.2\pm 0.943)\times 10^{-3}$ & $(8.05\pm 0.715)\times 10^{-3}$ \\
         \hline
         $\Delta t = 0.020$ & $(14.1\pm 1.20)\times 10^{-3}$ & $(11.5\pm 0.976)\times 10^{-3}$ & $(8.36\pm 0.775)\times 10^{-3}$ \\
         \hline
    \end{tabular}
    \caption{The dynamical factors ($F_{\rm D}$'s) for different values of $L$ and $\Delta t$.}
    \label{tab:dynamical}
\end{table}

Finally, combining both the statistical and dynamical factors,  we can obtain the nucleation rates via Eq.~\eqref{eq:rate-full}. Using the results of $\Delta t=0.01$ for the dynamical factor, the nucleation rates for all three lattice sizes are
\begin{equation}
    \Gamma_{\rm L,eff} = \begin{cases}
    \;(L=9)\;: \quad (9.2 \pm 0.76)\times 10^{-5} \vspace{1mm}\\
    (L=10): \quad (6.9 \pm 0.54)\times 10^{-5} \vspace{1mm}\\
    (L=11): \quad (5.3 \pm 0.47)\times 10^{-5}
    \end{cases} ~,
\end{equation}
which has the unit of $m^{-4}$ ($m=1$ in the simulations). 
To make a ``fair'' comparison to the saddle-point approximation, we apply a hard cutoff at $\Lambda=1/a=1$~\cite{Alford:1993ph} on the loop-corrected potentials [see Eqs.~\eqref{eq:V1} and \eqref{eq:VT}] and obtain $\Gamma_{b,{\rm eff}}^{\rm cut} = 2.22\times10^{-4}$ based on \texttt{CosmoTransitions} and \texttt{BubbleDet}. While this matches the lattice-simulated results at the order-of-magnitude level, we remark that a rigorous comparison should be made by taking the infinite-volume and continuum limits on the lattice results and leave this type of exercises to our future studies for $(3+1)$D models.

Before concluding this section, we briefly compare in more detail the semi-classical approach and the lattice formalism. In the semi-classical approach, the solution is based on the dominance of the bounce solution in the classical limit, which can be further corrected by including the higher-order corrections that consider both the spherical and non-spherical perturbations to the bubble profile. While in principle one can calculate the solution to arbitrary orders, the associated numerical complexity is considerably high; for more complicated models, even the one-loop correction is difficult to obtain (see the discussion in Ref.~\cite{Gould:2024chm}). On the other hand, the lattice formalism provides a nonperturbative way to directly investigate all the bubble solutions, at the cost of heavy simulations and the need for proper extrapolations. Ideally, at least for perturbative cases, one should expect that the two approaches would agree with each other when the semi-classical calculation is carried out to a high enough order and the continuum and infinite-volume limits are taken appropriately for the lattice formalism. However, not only is this hard to verify given the complexity of the necessary calculations, but also, as pointed out in Ref.~\cite{Gould:2024chm}, it remains an open question whether the higher-order corrections can actually make up the discrepancy. Therefore, the ``consistency'' between the semi-classical and lattice methods is something still to be verified in the future. 

\section{Discussion and conclusions}\label{sec:conclusions}

As briefly mentioned, a significant issue when increasing the lattice volume for bubble simulation is the dilution of the bubble configurations of interest due to bulk fluctuations. Specifically, the fluctuations of $\langle\phi\rangle$ in the purely metastable or stable phase scale roughly as $1/\sqrt{V}$, while the contribution of critical bubbles to $\langle\phi\rangle$ scales as $1/V$ due to their fixed size nature~\cite{Moore:2000jw}. Therefore, attempting to measure and extract bubble properties in the large-volume regime using the measurable $\langle\phi\rangle$ is likely to result in numerous ``false bubbles". To extrapolate lattice results to the infinite-volume limit, a suitable measurable that is less affected by significant bulk noise is required. One such measurable is $\langle\phi^2\rangle - 2A\langle\phi\rangle$, where $A$ is a manually chosen parameter. This measurable has been recently utilized in Ref.~\cite{Gould:2024chm}.

Another fundamental challenge not just for the PFMCMC method but also for the general flow-based simulation framework is the high training cost of NFs as the problem scale increases. In the context of the scalar FOPT problem we study, this challenge arises from several aspects: 1) the necessity of a small enough lattice spacing; 2) the need for a sufficiently large lattice size; and 3) the dimensions of the problem. The scalability issue of NF training has been discussed in the context of lattice QCD in Ref.~\cite{Abbott:2022zsh}, which presents inconclusive findings regarding the scaling nature of the problem but suggests several potential directions for future improvement. For instance, one direct approach is parallelizing the algorithm. Alternatively, if obtaining a high-quality flow model proves challenging due to resource constraints, a hybrid approach combining flow-based and traditional Monte Carlo methods may alleviate the issue by reducing the reliance on the model's quality while still leveraging its modeling capabilities to some extent.

While this paper serves as a proof of concept for the PFMCMC method, the immediate next step is to apply this method to a (3+1)D theory, such as the $\phi^4$ theory examined in Ref.~\cite{Gould:2024chm}, or other phenomenologically interesting models. An intriguing avenue of investigation lies in studying the non-sphericity of cosmological FOPT bubbles. This analysis could shed light on single-bubble gravitational wave generation, as recently explored in Ref.~\cite{Blum:2024hcs}. Additionally, numerically simulating bubbles and observing their evolution could improve our understanding of super-critical bubble expansion, including their interactions with the plasma~\cite{Bodeker:2017cim,Krajewski:2024gma}. These insights may contribute to more precise calculations of stochastic gravitational wave production from cosmological FOPTs and their detectability in gravitational wave experiments~\cite{Athron:2023xlk}.

In conclusion, we have demonstrated a novel flow-based algorithm, the PFMCMC method, capable of simulating cosmological FOPTs nonperturbatively, drawing on the nucleation framework outlined by Langer. Tested on various toy 2D models and the finite-temperature (2+1)D $\mathbb{Z}_2$-symmetric real scalar theory, our method yields results comparable to those found in the literature, effectively addressing highly hierarchical probability distribution challenges. Furthermore, we have illustrated how these results enable the calculation of the dynamical factor and nucleation rate of a thermal scalar field system. By integrating the power of NFs into the study of cosmological phase transitions, this method offers a promising approach for investigating higher-dimensional or more complex and realistic problems.

\vspace{1cm}
\subsubsection*{Acknowledgments}
The authors thank Kyle Cranmer for useful discussion. The work is supported by the U.S. Department of Energy under the contract DE-SC-0017647.  We thank the Center for High Throughput Computing at the University of Wisconsin-Madison (\url{https://chtc.cs.wisc.edu})~\cite{https://doi.org/10.21231/gnt1-hw21} for providing the computing resources.

\appendix

\section{The nucleation theory for relativistic quantum fields}
\label{sec:app-nucleation-QFT}

In this appendix, we review the nucleation theory for relativistic quantum fields, first derived by Coleman for zero-temperature theories~\cite{Coleman:1977py} and later generalized for finite-temperature theories by Linde~\cite{Linde:1980tt,Linde:1981zj}. In Sec.~\ref{sec:nucleation}, we reviewed how the nucleation rate for a classical field can be decomposed into a dynamical factor $F_{\rm D}$ and a statistical factor $F_{\rm S}$ [see Eq.~\eqref{eq:rate-full}]. By matching the free energy functional of the classical field $F[\phi]/T$ to the Euclidean quantum field action functional $S_{\rm E}[\phi]$, $F_{\rm S}$ can be calculated from the fluctuations around the saddle point and is given up to the second order by~\cite{Coleman:1977py,Callan:1977pt,Vainshtein:1981wh,Andreassen:2016cvx,Andreassen:2017rzq,Ai:2019fri}
\begin{equation}
    F_{\rm S} = \left(\frac{S_{\rm E}[\phi_b]}{2\pi}\right)^{(d-1)/2}\left\vert\dfrac{\det\big(-\nabla^2+V^{\prime\prime}(\phi_F)\big)}{\det^\prime\big(-\nabla^2+V^{\prime\prime}(\phi_b)\big)}\right\vert^{1/2} e^{-(S_{\rm E}[\phi_b]-S_{\rm E}[\phi_F])} ~,
\end{equation}
where $S_{\rm E}$ is the finite-temperature Euclidean action of the system and given by
\begin{equation}\label{eq:action}
    S_{\rm E}[\phi] = \frac{1}{T}\int d^{d-1}x\left[-\frac{1}{2}\phi\nabla^2\phi + V(\phi)\right] \equiv \frac{S[\phi]}{T} ~.
\end{equation}
Here, $d$ is the spacetime dimension; $\phi_F$ is the metastable vacuum configuration; $\det^\prime$ signifies that the zero modes are omitted from the functional determinant. Assuming spherical symmetry for the system, $\phi_b$ is solved using the equation of motion~\cite{Coleman:1977py}
\begin{equation}
    \left.\frac{\delta S[\phi]}{\delta\phi}\right\vert_{\phi_b} = -\partial_r^2\phi_b(r) - \frac{d-2}{r}\partial_r\phi_b(r) + V^\prime(\phi_b) = 0 ~,
\end{equation}
subject to the boundary conditions
\begin{equation}
    \lim_{r\to\infty}\phi_b(r) = \phi_F ,~ \left.\partial_r\phi_b\right\vert_{r=0}=0 ~.
\end{equation}
On the other hand, for a real scalar theory, $F_{\rm D}$ can be calculated using Langevin dynamics and is given by~\cite{Langer:1969bc,Ai:2018guc,Chigusa:2020jbn}
\begin{equation}
    F_{\rm D} = \frac{1}{2\pi}\left(\sqrt{\vert\lambda_-\vert+\frac{\tilde{\eta}^2}{4}}-\frac{\tilde{\eta}}{2}\right) ~,
\end{equation}
where $\lambda_-$ is the negative eigenvalue of the functional second derivative $-\nabla^2+V^{\prime\prime}(\phi_b)$ and $\tilde{\eta}$ is the Langevin damping coefficient, which has to be determined with real-time input.

\section{Finite-temperature (2+1)D $\mathbb{Z}_2$-symmetric scalar theory}\label{sec:app-finite-temp}

In this appendix, we study the finite-temperature (2+1)D $\mathbb{Z}_2$-symmetric scalar theory. The most general renormalizable tree-level potential is given by
\begin{equation}\label{eq:thermal:V0}
    V_0(\phi) = \frac{m^2}{2}\phi^2 + \frac{\lambda}{4!}\phi^4 + \frac{\eta}{6!}\phi^6 ~.
\end{equation}
In order to calculate the thermal effective mass, we consider the self-energy corrections to $\phi$ up to two-loop since the six-point interaction in Eq.~\eqref{eq:thermal:V0} will contribute at the leading perturbative order. Diagrammatically, this is given by 
\begin{equation}\label{eq:meff}
\begin{aligned}
    m^2_{\rm eff,0}(\phi,T) &= \begin{tikzpicture}[baseline=(i.base)]
	\begin{feynman}
		\vertex (i);
		\vertex[crossed dot] at ($(i) + (1.5cm, 0.0cm)$) (a){};
		\vertex at ($(a) + (1.5cm, 0.0cm)$) (o);
	
	\diagram* {
		(i) -- (a) -- (o);
	};
	\end{feynman}
	\end{tikzpicture} + \begin{tikzpicture}[baseline=(i.base)]
	\begin{feynman}
		\vertex (i);
		\vertex at ($(i) + (1.5cm, 0.0cm)$) (a);
		\vertex at ($(a) + (1.5cm, 0.0cm)$) (o);
		\vertex at ($(a) + (0.3cm, 0.9cm)$) (d1);
		\vertex at ($(a) + (-0.3cm, 0.9cm)$) (d2); 
	
	\diagram* {
		(i) -- (a) -- (o);
		(d1) -- (a) -- (d2);	
	};
	\end{feynman}
	\end{tikzpicture} + \begin{tikzpicture}[baseline=(i.base)]
	\begin{feynman}
		\vertex (i);
		\vertex at ($(i) + (1.5cm, 0.0cm)$) (a);
		\vertex at ($(a) + (1.5cm, 0.0cm)$) (o);
		\vertex at ($(a) + (0.3cm, 0.9cm)$) (d1);
		\vertex at ($(a) + (-0.3cm, 0.9cm)$) (d2); 
		\vertex at ($(a) + (0.3cm, -0.9cm)$) (d3);
		\vertex at ($(a) + (-0.3cm, -0.9cm)$) (d4); 
	
	\diagram* {
		(i) -- (a) -- (o);
		(d1) -- (a) -- (d2);
		(d3) -- (a) -- (d4);
	};
	\end{feynman}
	\end{tikzpicture} \\
	&\quad + \begin{tikzpicture}[baseline=(i.base)]
	\begin{feynman}
		\vertex (i);
		\vertex at ($(i) + (1.5cm, 0.0cm)$) (a);
		\vertex at ($(a) + (1.5cm, 0.0cm)$) (o);
		\vertex at ($(a) + (0.0cm, 0.9cm)$) (b);
	
	\diagram* {
		(i) -- (a) -- (o);
		(a) --[out=70, in=-5] (b) --[out=185, in=110] (a);	
	};
	\end{feynman}
	\end{tikzpicture} + \begin{tikzpicture}[baseline=(i.base)]
	\begin{feynman}
		\vertex (i);
		\vertex at ($(i) + (1.5cm, 0.0cm)$) (a);
		\vertex at ($(a) + (1.5cm, 0.0cm)$) (o);
		\vertex at ($(a) + (0.0cm, 0.9cm)$) (b);
		\vertex at ($(a) + (0.3cm, -0.9cm)$) (d1);
		\vertex at ($(a) + (-0.3cm, -0.9cm)$) (d2); 
	
	\diagram* {
		(i) -- (a) -- (o);
		(a) --[out=70, in=-5] (b) --[out=185, in=110] (a);
		(d1) -- (a) -- (d2);
	};
	\end{feynman}
	\end{tikzpicture} + \begin{tikzpicture}[baseline=(i.base)]
	\begin{feynman}
		\vertex (i);
		\vertex at ($(i) + (1.5cm, 0.0cm)$) (a);
		\vertex at ($(a) + (1.5cm, 0.0cm)$) (o);
		\vertex at ($(a) + (0.0cm, 0.9cm)$) (b);
		\vertex at ($(a) + (0.0cm, -0.9cm)$) (c);
	
	\diagram* {
		(i) -- (a) -- (o);
		(a) --[out=70, in=-5] (b) --[out=185, in=110] (a);
		(a) --[out=-70, in=5] (c) --[out=-185, in=-110] (a);
	};
	\end{feynman}
	\end{tikzpicture} \\
    &= m^2 + \frac{\lambda}{2}\phi^2 + \frac{\eta}{4!}\phi^4 + \frac{\lambda}{2}\tau_1(T,m) + \frac{\eta}{4}\phi^2\tau_1(T,m) + \frac{\eta}{8}\tau_1^2(T,m) ~,
\end{aligned}
\end{equation}
where $\tau_1(T,m)$ is the thermal-loop bubble, as given by
\begin{equation}
	\tau_1(T,m) = \begin{tikzpicture}[baseline=(l.base)]
	\begin{feynman}
		\vertex (l);
		\vertex at ($(l) + (1.0cm, 0.0cm)$) (r);
		\vertex at ($(l)!0.5!(r) + (0.0cm, 0.5cm)$) (t);
		\vertex at ($(l)!0.5!(r) + (0.0cm, -0.5cm)$) (b);
	\diagram* {
		(l) --[out=90, in=180] (t) --[out=0, in=90] (r) --[out=-90, in=0] (b) --[out=180, in=-90] (l);
	};
	\end{feynman}
	\end{tikzpicture} = -\frac{T}{2\pi}\ln\left(2\sinh\frac{m}{2T}\right) ~.
\end{equation}
As pointed out in Ref.~\cite{Appelquist:1981sf,Dilkes:1999uf}, $\tau_1$ is manifestly IR-divergent in the $m/T\to0$ limit. Therefore, we introduce the thermal mass (Debye mass) $m_T$ as the IR-cutoff, which can be obtained through solving the gap equation
\begin{equation}\label{eq:mT}
\begin{aligned}
	m_T^2 &= \begin{tikzpicture}[baseline=(i.base)]
	\begin{feynman}
		\vertex (i);
		\vertex[crossed dot] at ($(i) + (1.5cm, 0.0cm)$) (a){};
		\vertex at ($(a) + (1.5cm, 0.0cm)$) (o);
	
	\diagram* {
		(i) -- (a) -- (o);
	};
	\end{feynman}
	\end{tikzpicture} + \begin{tikzpicture}[baseline=(i.base)]
	\begin{feynman}
		\vertex (i);
		\vertex at ($(i) + (1.5cm, 0.0cm)$) (a);
		\vertex at ($(a) + (1.5cm, 0.0cm)$) (o);
		\vertex at ($(a) + (0.0cm, 0.9cm)$) (b);
	
	\diagram* {
		(i) -- (a) -- (o);
		(a) --[out=70, in=-5] (b) --[out=185, in=110] (a);	
	};
	\end{feynman}
	\end{tikzpicture} + \begin{tikzpicture}[baseline=(i.base)]
	\begin{feynman}
		\vertex (i);
		\vertex at ($(i) + (1.5cm, 0.0cm)$) (a);
		\vertex at ($(a) + (1.5cm, 0.0cm)$) (o);
		\vertex at ($(a) + (0.0cm, 0.9cm)$) (b);
		\vertex at ($(a) + (0.0cm, -0.9cm)$) (c);
	
	\diagram* {
		(i) -- (a) -- (o);
		(a) --[out=70, in=-5] (b) --[out=185, in=110] (a);
		(a) --[out=-70, in=5] (c) --[out=-185, in=-110] (a);
	};
	\end{feynman}
	\end{tikzpicture} \\
        &= m^2 + \frac{\lambda}{2}\tau_1(T,m_T) + \frac{\eta}{8}\tau_1^2(T,m_T) ~.
\end{aligned}
\end{equation} 
Consequently, $m_{\rm eff}$ is given by
\begin{equation}
    m_{\rm eff}^2(\phi,T) = m_T^2 + \frac{\lambda}{2}\phi^2 + \frac{\eta}{4!}\phi^4 + \frac{\eta}{4}\phi^2\tau_1(T,m_T) ~,
\end{equation}
in contrast to the effective mass-squared $m_{\rm eff,0}^2$ without IR-regularization given in Eq.~\eqref{eq:meff}.

For a generic theory, one can in principle separate the loop corrections to the effective potential into two parts: the zero-temperature Coleman-Weinberg potential $V_1$ and the finite-temperature effective potential $V_\beta$. However, because $m_T$ is required for IR regularization in this model, the usual Coleman-Weinberg potential becomes temperature-dependent, $V_1=V_1(\phi,T)$. Diagrammatically, these two corrections are given by
\begin{equation}
\begin{aligned}
    &V_1(\phi,T)+V_\beta(\phi,T) \\ 
    &\quad = \begin{tikzpicture}[baseline=(a.base)]
    \begin{feynman}
        \vertex (a);
        \vertex at ($(a) + (1.5cm, 0.0cm)$) (b);
        \vertex at ($(a) + (-0.5cm, 0.5cm)$) (i1);
        \vertex at ($(a) + (-0.5cm, -0.5cm)$) (i2);

        \diagram* {
            (a) --[out=80, in=100] (b) --[out=260, in=280] (a);
            (a) -- (i1);
            (a) -- (i2);
        };
    \end{feynman}
    \end{tikzpicture} + \begin{tikzpicture}[baseline=(a.base)]
    \begin{feynman}
        \vertex (a);
        \vertex at ($(a) + (1.5cm, 0.0cm)$) (b);
        \vertex at ($(a) + (-0.5cm, 0.5cm)$) (i1);
        \vertex at ($(a) + (-0.5cm, -0.5cm)$) (i2);
        \vertex at ($(a) + (-0.5cm, 0.15cm)$) (i3);
        \vertex at ($(a) + (-0.5cm, -0.15cm)$) (i4);

        \diagram* {
            (a) --[out=80, in=100] (b) --[out=260, in=280] (a);
            (a) -- (i1);
            (a) -- (i2);
            (a) -- (i3);
            (a) -- (i4);
        };
    \end{feynman}
    \end{tikzpicture} + \begin{tikzpicture}[baseline=(a.base)]
    \begin{feynman}
        \vertex (a);
        \vertex at ($(a) + (1.5cm, 0.0cm)$) (b);
        \vertex at ($(a) + (-0.5cm, 0.5cm)$) (i1);
        \vertex at ($(a) + (-0.5cm, -0.5cm)$) (i2);
        \vertex at ($(b) + (0.5cm, 0.5cm)$) (o1);
        \vertex at ($(b) + (0.5cm, -0.5cm)$) (o2);

        \diagram* {
            (a) --[out=80, in=100] (b) --[out=260, in=280] (a);
            (a) -- (i1);
            (a) -- (i2);
            (b) -- (o1);
            (b) -- (o2);
        };
    \end{feynman}
    \end{tikzpicture} + \begin{tikzpicture}[baseline=(a.base)]
    \begin{feynman}
        \vertex (a);
        \vertex at ($(a) + (1.5cm, 0.0cm)$) (b);
        \vertex at ($(a) + (-0.5cm, 0.5cm)$) (i1);
        \vertex at ($(a) + (-0.5cm, -0.5cm)$) (i2);
        \vertex at ($(a) + (-0.5cm, 0.15cm)$) (i3);
        \vertex at ($(a) + (-0.5cm, -0.15cm)$) (i4);
        \vertex at ($(b) + (0.5cm, 0.5cm)$) (o1);
        \vertex at ($(b) + (0.5cm, -0.5cm)$) (o2);

        \diagram* {
            (a) --[out=80, in=100] (b) --[out=260, in=280] (a);
            (a) -- (i1);
            (a) -- (i2);
            (a) -- (i3);
            (a) -- (i4);
            (b) -- (o1);
            (b) -- (o2);
        };
    \end{feynman}
    \end{tikzpicture} + \begin{tikzpicture}[baseline=(a.base)]
    \begin{feynman}
        \vertex (a);
        \vertex at ($(a) + (1.5cm, 0.0cm)$) (b);
        \vertex at ($(a) + (-0.5cm, 0.5cm)$) (i1);
        \vertex at ($(a) + (-0.5cm, -0.5cm)$) (i2);
        \vertex at ($(a) + (-0.5cm, 0.15cm)$) (i3);
        \vertex at ($(a) + (-0.5cm, -0.15cm)$) (i4);
        \vertex at ($(b) + (0.5cm, 0.5cm)$) (o1);
        \vertex at ($(b) + (0.5cm, -0.5cm)$) (o2);
        \vertex at ($(b) + (0.5cm, 0.15cm)$) (o3);
        \vertex at ($(b) + (0.5cm, -0.15cm)$) (o4);

        \diagram* {
            (a) --[out=80, in=100] (b) --[out=260, in=280] (a);
            (a) -- (i1);
            (a) -- (i2);
            (a) -- (i3);
            (a) -- (i4);
            (b) -- (o1);
            (b) -- (o2);
            (b) -- (o3);
            (b) -- (o4);
        };
    \end{feynman}
    \end{tikzpicture} + \cdots ~,
\end{aligned}
\end{equation}
where
\begin{equation}\label{eq:V1}
    V_1(\phi,T) = \frac{1}{2}\int\frac{d^3p}{(2\pi)^3}\log\left[p^2+m_{\rm eff}^2(\phi,T)\right] = -\frac{1}{12\pi}m_{\rm eff}^3(\phi,T) ~,
\end{equation}
and
\begin{equation}\label{eq:VT}
    V_\beta(\phi,T) = T\int\frac{d^2p}{(2\pi)^2}\log\left[1-e^{-\frac{1}{T}\sqrt{p^2+m_{\rm eff}^2(\phi,T)}}\right] = -\frac{T^3}{2\pi}\left[\frac{m_{\rm eff}}{T}{\rm Li}_2\left(e^{-\frac{m_{\rm eff}}{T}}\right) + {\rm Li}_3\left(e^{-\frac{m_{\rm eff}}{T}}\right)\right] ~,
\end{equation}
where ${\rm Li}_n$ is the polylogarithmic function. Overall, the finite-temperature effective potential is given by
\beqa
    V_{\rm eff}(\phi,T) = V_0(\phi) + V_1(\phi,T) + V_\beta(\phi,T) ~.
\eeqa

We briefly discuss the perturbativity bounds on $\lambda$ and $\eta$. For $\lambda$, we consider the diagram
\begin{equation}
	I_\lambda = 
\begin{tikzpicture}[baseline=(a.base)]
\begin{feynman}
	\vertex (l1);
	\vertex at ($(l1) + (0.0cm, -1.0cm)$) (l3);
	\vertex at ($(l1) + (4.0cm, 0.0cm)$) (r1);
	\vertex at ($(l3) + (4.0cm, 0.0cm)$) (r3);
	\vertex at ($(l1)!0.5!(l3) + (1.0cm, 0.0cm)$) (a);
	\vertex at ($(r1)!0.5!(r3) + (-1.0cm, 0.0cm)$) (b);
	
	\diagram* {
		(l1) -- (a); (l3) -- (a);
		(r1) -- (b); (r3) -- (b);
		(a) --[out=85, in=95, momentum=$q$] (b);
		(a) --[out=-85, in=-95, rmomentum'=$q$] (b);
	};
\end{feynman}
\end{tikzpicture}
	\sim \frac{\lambda^2}{2}\int\frac{d^3q}{(2\pi)^3}\frac{1}{(q^2+m^2)^2} = \frac{\lambda^2}{2\times8\pi m} ~,
\end{equation}
where we ignore all the $\mathcal{O}(1)$ factors. By requiring $\vert I_\lambda\vert\lesssim\vert\lambda\vert$, the perturbativity bound on $\lambda$ is roughly given by $\vert\lambda/m\vert\lesssim 8\pi$. For $\eta$, we consider the diagram 
\begin{equation}
	I_\eta^{(2)} =
\begin{tikzpicture}[baseline=(a.base)]
\begin{feynman}
	\vertex (l1);
	\vertex at ($(l1) + (0.0cm, -0.5cm)$) (l2);
	\vertex at ($(l2) + (0.0cm, -0.5cm)$) (l3);
	\vertex at ($(l1) + (4.0cm, 0.0cm)$) (r1);
	\vertex at ($(l2) + (4.0cm, 0.0cm)$) (r2);
	\vertex at ($(l3) + (4.0cm, 0.0cm)$) (r3);
	\vertex at ($(l2) + (1.0cm, 0.0cm)$) (a);
	\vertex at ($(r2) + (-1.0cm, 0.0cm)$) (b);
	
	\diagram* {
		(l1) -- (a); (l2) -- (a); (l3) -- (a);
		(r1) -- (b); (r2) -- (b); (r3) -- (b);
		(a) -- (b);
		(a) --[out=85, in=95, momentum=$q_1$] (b);
		(a) --[out=-85, in=-95, momentum'=$q_2$] (b);
	};
\end{feynman}
\end{tikzpicture} 
        \sim \frac{\eta^2}{3}\int \frac{d^3q_1d^3q_2}{(2\pi)^6}\frac{1}{q_1^2q_2^2(q_1+q_2)^2} \sim \frac{\eta^2}{3\times(2\pi^2)^2}\int \frac{dq}{q}  ~,
\end{equation}
where we ignore all the $\mathcal{O}(1)$ factors. By requiring $\vert I_\eta^{(2)}\vert\lesssim\vert \eta\vert$ (or the tree-level coupling), the perturbativity bound on $\eta$ is roughly given by $\vert\eta\vert\lesssim (2\pi^2)^2$.

Before concluding, we remark that the non-analytic structure of $V_1(\phi,T)$ and $V_\beta(\phi,T)$ can possibly induce imaginary components, which suggests the instability of the homogeneous field state as pointed out in Refs.~\cite{Weinberg:1987vp,Sher:1988mj}. Nevertheless, if the imaginary part of $V_{\rm eff}$ is sufficiently small compared to its real part, the decay timescale is then negligible compared to the transition timescale. In this case, we can safely extract the real part and use it to perform phase transition and nucleation calculations.

\setlength{\bibsep}{3pt}
\providecommand{\href}[2]{#2}\begingroup\raggedright\endgroup


\begin{thebibliography}{10}

\bibitem{Kuzmin:1985mm}
V.~A. Kuzmin, V.~A. Rubakov, and M.~E. Shaposhnikov, {\it {On the Anomalous
  Electroweak Baryon Number Nonconservation in the Early Universe}},  {\em
  Phys. Lett. B} {\bf 155} (1985) 36.

\bibitem{Cohen:1993nk}
A.~G. Cohen, D.~B. Kaplan, and A.~E. Nelson, {\it {Progress in electroweak
  baryogenesis}},  {\em Ann. Rev. Nucl. Part. Sci.} {\bf 43} (1993) 27--70,
  [\href{http://arxiv.org/abs/hep-ph/9302210}{{\tt hep-ph/9302210}}].

\bibitem{Bai:2023cqj}
Y.~Bai, T.-K. Chen, and M.~Korwar, {\it {QCD-collapsed domain walls: QCD phase
  transition and gravitational wave spectroscopy}},  {\em JHEP} {\bf 12} (2023)
  194, [\href{http://arxiv.org/abs/2306.17160}{{\tt arXiv:2306.17160}}].

\bibitem{NANOGrav:2020bcs}
{\bf NANOGrav} Collaboration, Z.~Arzoumanian et~al., {\it {The NANOGrav 12.5 yr
  Data Set: Search for an Isotropic Stochastic Gravitational-wave Background}},
   {\em Astrophys. J. Lett.} {\bf 905} (2020), no.~2 L34,
  [\href{http://arxiv.org/abs/2009.04496}{{\tt arXiv:2009.04496}}].

\bibitem{EPTA:2021crs}
{\bf EPTA} Collaboration, S.~Chen et~al., {\it {Common-red-signal analysis with
  24-yr high-precision timing of the European Pulsar Timing Array: inferences
  in the stochastic gravitational-wave background search}},  {\em Mon. Not.
  Roy. Astron. Soc.} {\bf 508} (2021), no.~4 4970--4993,
  [\href{http://arxiv.org/abs/2110.13184}{{\tt arXiv:2110.13184}}].

\bibitem{Goncharov:2021oub}
B.~Goncharov et~al., {\it {On the Evidence for a Common-spectrum Process in the
  Search for the Nanohertz Gravitational-wave Background with the Parkes Pulsar
  Timing Array}},  {\em Astrophys. J. Lett.} {\bf 917} (2021), no.~2 L19,
  [\href{http://arxiv.org/abs/2107.12112}{{\tt arXiv:2107.12112}}].

\bibitem{Perera:2019sca}
B.~B.~P. Perera et~al., {\it {The International Pulsar Timing Array: Second
  data release}},  {\em Mon. Not. Roy. Astron. Soc.} {\bf 490} (2019), no.~4
  4666--4687, [\href{http://arxiv.org/abs/1909.04534}{{\tt arXiv:1909.04534}}].

\bibitem{Athron:2023xlk}
P.~Athron, C.~Bal\'azs, A.~Fowlie, L.~Morris, and L.~Wu, {\it {Cosmological
  phase transitions: From perturbative particle physics to gravitational
  waves}},  {\em Prog. Part. Nucl. Phys.} {\bf 135} (2024) 104094,
  [\href{http://arxiv.org/abs/2305.02357}{{\tt arXiv:2305.02357}}].

\bibitem{Langer:1969bc}
J.~S. Langer, {\it {Statistical theory of the decay of metastable states}},
  {\em Annals Phys.} {\bf 54} (1969) 258--275.

\bibitem{Coleman:1977py}
S.~R. Coleman, {\it {The Fate of the False Vacuum. 1. Semiclassical Theory}},
  {\em Phys. Rev. D} {\bf 15} (1977) 2929--2936. [Erratum: Phys.Rev.D 16, 1248
  (1977)].

\bibitem{Linde:1980tt}
A.~D. Linde, {\it {Fate of the False Vacuum at Finite Temperature: Theory and
  Applications}},  {\em Phys. Lett. B} {\bf 100} (1981) 37--40.

\bibitem{Linde:1981zj}
A.~D. Linde, {\it {Decay of the False Vacuum at Finite Temperature}},  {\em
  Nucl. Phys. B} {\bf 216} (1983) 421. [Erratum: Nucl.Phys.B 223, 544 (1983)].

\bibitem{Moore:2000jw}
G.~D. Moore and K.~Rummukainen, {\it {Electroweak bubble nucleation,
  nonperturbatively}},  {\em Phys. Rev. D} {\bf 63} (2001) 045002,
  [\href{http://arxiv.org/abs/hep-ph/0009132}{{\tt hep-ph/0009132}}].

\bibitem{Gould:2022ran}
O.~Gould, S.~G\"uyer, and K.~Rummukainen, {\it {First-order electroweak phase
  transitions: A nonperturbative update}},  {\em Phys. Rev. D} {\bf 106}
  (2022), no.~11 114507, [\href{http://arxiv.org/abs/2205.07238}{{\tt
  arXiv:2205.07238}}].

\bibitem{Gould:2024chm}
O.~Gould, A.~Kormu, and D.~J. Weir, {\it {A nonperturbative test of nucleation
  calculations for strong phase transitions}},
  \href{http://arxiv.org/abs/2404.01876}{{\tt arXiv:2404.01876}}.

\bibitem{Wolff:1989wq}
U.~Wolff, {\it {Critical Slowing Down}},  {\em Nucl. Phys. B Proc. Suppl.} {\bf
  17} (1990) 93--102.

\bibitem{DelDebbio:2004xh}
L.~Del~Debbio, G.~M. Manca, and E.~Vicari, {\it {Critical slowing down of
  topological modes}},  {\em Phys. Lett. B} {\bf 594} (2004) 315--323,
  [\href{http://arxiv.org/abs/hep-lat/0403001}{{\tt hep-lat/0403001}}].

\bibitem{Meyer:2006ty}
H.~B. Meyer, H.~Simma, R.~Sommer, M.~Della~Morte, O.~Witzel, and U.~Wolff, {\it
  {Exploring the HMC trajectory-length dependence of autocorrelation times in
  lattice QCD}},  {\em Comput. Phys. Commun.} {\bf 176} (2007) 91--97,
  [\href{http://arxiv.org/abs/hep-lat/0606004}{{\tt hep-lat/0606004}}].

\bibitem{Schaefer:2009xx}
S.~Schaefer, R.~Sommer, and F.~Virotta, {\it {Investigating the critical
  slowing down of QCD simulations}},  {\em PoS} {\bf LAT2009} (2009) 032,
  [\href{http://arxiv.org/abs/0910.1465}{{\tt arXiv:0910.1465}}].

\bibitem{Schaefer:2010hu}
{\bf ALPHA} Collaboration, S.~Schaefer, R.~Sommer, and F.~Virotta, {\it
  {Critical slowing down and error analysis in lattice QCD simulations}},  {\em
  Nucl. Phys. B} {\bf 845} (2011) 93--119,
  [\href{http://arxiv.org/abs/1009.5228}{{\tt arXiv:1009.5228}}].

\bibitem{Hackett:2021idh}
D.~C. Hackett, C.-C. Hsieh, M.~S. Albergo, D.~Boyda, J.-W. Chen, K.-F. Chen,
  K.~Cranmer, G.~Kanwar, and P.~E. Shanahan, {\it {Flow-based sampling for
  multimodal distributions in lattice field theory}},
  \href{http://arxiv.org/abs/2107.00734}{{\tt arXiv:2107.00734}}.

\bibitem{Nicoli:2023qsl}
K.~A. Nicoli, C.~J. Anders, T.~Hartung, K.~Jansen, P.~Kessel, and S.~Nakajima,
  {\it {Detecting and mitigating mode-collapse for flow-based sampling of
  lattice field theories}},  {\em Phys. Rev. D} {\bf 108} (2023), no.~11
  114501, [\href{http://arxiv.org/abs/2302.14082}{{\tt arXiv:2302.14082}}].

\bibitem{gao2023rare}
Z.~Gao, D.~Zhang, L.~Daniel, and D.~S. Boning, {\it Rare event probability
  learning by normalizing flows},  \href{http://arxiv.org/abs/2310.19167}{{\tt
  arXiv:2310.19167}}.

\bibitem{Albergo:2019eim}
M.~S. Albergo, G.~Kanwar, and P.~E. Shanahan, {\it {Flow-based generative
  models for Markov chain Monte Carlo in lattice field theory}},  {\em Phys.
  Rev. D} {\bf 100} (2019), no.~3 034515,
  [\href{http://arxiv.org/abs/1904.12072}{{\tt arXiv:1904.12072}}].

\bibitem{Berg:1992qua}
B.~A. Berg and T.~Neuhaus, {\it {Multicanonical ensemble: A New approach to
  simulate first order phase transitions}},  {\em Phys. Rev. Lett.} {\bf 68}
  (1992) 9--12, [\href{http://arxiv.org/abs/hep-lat/9202004}{{\tt
  hep-lat/9202004}}].

\bibitem{10.1214/aos/1176325750}
L.~Tierney, {\it {Markov Chains for Exploring Posterior Distributions}},  {\em
  The Annals of Statistics} {\bf 22} (1994), no.~4 1701 -- 1728.

\bibitem{Chien_2014}
C.-C. Chien, J.-H. She, and F.~Cooper, {\it Mean-field description of pairing
  effects, bkt physics, and superfluidity in 2d bose gases},  {\em Annals of
  Physics} {\bf 347} (Aug., 2014) 192–206,
  [\href{http://arxiv.org/abs/1203.3254}{{\tt arXiv:1203.3254}}].

\bibitem{Konietin_2017}
P.~Konietin and V.~Pastukhov, {\it 2d dilute bose mixture at low temperatures},
   {\em Journal of Low Temperature Physics} {\bf 190} (Dec., 2017) 256–266,
  [\href{http://arxiv.org/abs/1708.00432}{{\tt arXiv:1708.00432}}].

\bibitem{Pastukhov_2018}
V.~Pastukhov, {\it Polaron in dilute 2d bose gas at low temperatures},  {\em
  Journal of Physics B: Atomic, Molecular and Optical Physics} {\bf 51} (July,
  2018) 155203, [\href{http://arxiv.org/abs/1712.06978}{{\tt
  arXiv:1712.06978}}].

\bibitem{Pastukhov_2018_2}
V.~Pastukhov, {\it Ground-state properties of a dilute two-dimensional bose
  gas},  {\em Journal of Low Temperature Physics} {\bf 194} (Oct., 2018)
  197–208, [\href{http://arxiv.org/abs/1803.05242}{{\tt arXiv:1803.05242}}].

\bibitem{Csernai:1992tj}
L.~P. Csernai and J.~I. Kapusta, {\it {Nucleation of relativistic first order
  phase transitions}},  {\em Phys. Rev. D} {\bf 46} (1992) 1379--1390.

\bibitem{Moore:2001vf}
G.~D. Moore, K.~Rummukainen, and A.~Tranberg, {\it {Nonperturbative computation
  of the bubble nucleation rate in the cubic anisotropy model}},  {\em JHEP}
  {\bf 04} (2001) 017, [\href{http://arxiv.org/abs/hep-lat/0103036}{{\tt
  hep-lat/0103036}}].

\bibitem{Goldenfeld:1992qy}
N.~Goldenfeld, {\em Lectures On Phase Transitions And The Renormalization
  Group}.
\newblock Frontiers in physics. Westview Press, 1992.

\bibitem{Moore:2000mx}
G.~D. Moore, {\it {Sphaleron rate in the symmetric electroweak phase}},  {\em
  Phys. Rev. D} {\bf 62} (2000) 085011,
  [\href{http://arxiv.org/abs/hep-ph/0001216}{{\tt hep-ph/0001216}}].

\bibitem{critical-pheno}
J.~Zinn-Justin, {\em {Quantum Field Theory and Critical Phenomena: Fifth
  Edition}}.
\newblock Oxford University Press, 04, 2021.

\bibitem{mode-seeking}
T.~Minka et~al., {\it Divergence measures and message passing},  tech. rep.,
  Microsoft Research, 2005.

\bibitem{Kobyzev_2021}
I.~Kobyzev, S.~J. Prince, and M.~A. Brubaker, {\it Normalizing flows: An
  introduction and review of current methods},  {\em IEEE Transactions on
  Pattern Analysis and Machine Intelligence} {\bf 43} (Nov., 2021) 3964–3979.

\bibitem{dinh2017density}
L.~Dinh, J.~Sohl-Dickstein, and S.~Bengio, {\it Density estimation using real
  nvp},  \href{http://arxiv.org/abs/1605.08803}{{\tt arXiv:1605.08803}}.

\bibitem{KL-divergence}
S.~Kullback and R.~A. Leibler, {\it {On Information and Sufficiency}},  {\em
  The Annals of Mathematical Statistics} {\bf 22} (1951), no.~1 79 -- 86.

\bibitem{DelDebbio:2021qwf}
L.~Del~Debbio, J.~M. Rossney, and M.~Wilson, {\it {Efficient modeling of
  trivializing maps for lattice \ensuremath{\phi}4 theory using normalizing
  flows: A first look at scalability}},  {\em Phys. Rev. D} {\bf 104} (2021),
  no.~9 094507, [\href{http://arxiv.org/abs/2105.12481}{{\tt
  arXiv:2105.12481}}].

\bibitem{Shen:2022xcm}
J.~Shen, P.~Draper, and A.~X. El-Khadra, {\it {Vacuum decay and Euclidean
  lattice Monte~Carlo}},  {\em Phys. Rev. D} {\bf 107} (2023), no.~9 094506,
  [\href{http://arxiv.org/abs/2210.05925}{{\tt arXiv:2210.05925}}].

\bibitem{ziegler2019latent}
Z.~M. Ziegler and A.~M. Rush, {\it Latent normalizing flows for discrete
  sequences},  \href{http://arxiv.org/abs/1901.10548}{{\tt arXiv:1901.10548}}.

\bibitem{Wainwright:2011kj}
C.~L. Wainwright, {\it {CosmoTransitions: Computing Cosmological Phase
  Transition Temperatures and Bubble Profiles with Multiple Fields}},  {\em
  Comput. Phys. Commun.} {\bf 183} (2012) 2006--2013,
  [\href{http://arxiv.org/abs/1109.4189}{{\tt arXiv:1109.4189}}].

\bibitem{Ekstedt:2023sqc}
A.~Ekstedt, O.~Gould, and J.~Hirvonen, {\it {BubbleDet: a Python package to
  compute functional determinants for bubble nucleation}},  {\em JHEP} {\bf 12}
  (2023) 056, [\href{http://arxiv.org/abs/2308.15652}{{\tt arXiv:2308.15652}}].

\bibitem{Gould:2021dzl}
O.~Gould, {\it {Real scalar phase transitions: a nonperturbative analysis}},
  {\em JHEP} {\bf 04} (2021) 057, [\href{http://arxiv.org/abs/2101.05528}{{\tt
  arXiv:2101.05528}}].

\bibitem{Alford:1993ph}
M.~G. Alford and M.~Gleiser, {\it {Metastability in two-dimensions and the
  effective potential}},  {\em Phys. Rev. D} {\bf 48} (1993) 2838--2844,
  [\href{http://arxiv.org/abs/hep-ph/9304245}{{\tt hep-ph/9304245}}].

\bibitem{Abbott:2022zsh}
R.~Abbott et~al., {\it {Aspects of scaling and scalability for flow-based
  sampling of lattice QCD}},  {\em Eur. Phys. J. A} {\bf 59} (2023), no.~11
  257, [\href{http://arxiv.org/abs/2211.07541}{{\tt arXiv:2211.07541}}].

\bibitem{Blum:2024hcs}
K.~Blum and M.~Mirbabayi, {\it {A single-bubble source for gravitational waves
  in a cosmological phase transition}},
  \href{http://arxiv.org/abs/2403.20164}{{\tt arXiv:2403.20164}}.

\bibitem{Bodeker:2017cim}
D.~Bodeker and G.~D. Moore, {\it {Electroweak Bubble Wall Speed Limit}},  {\em
  JCAP} {\bf 05} (2017) 025, [\href{http://arxiv.org/abs/1703.08215}{{\tt
  arXiv:1703.08215}}].

\bibitem{Krajewski:2024gma}
T.~Krajewski, M.~Lewicki, and M.~Zych, {\it {Bubble-wall velocity in local
  thermal equilibrium: hydrodynamical simulations vs analytical treatment}},
  \href{http://arxiv.org/abs/2402.15408}{{\tt arXiv:2402.15408}}.

\bibitem{https://doi.org/10.21231/gnt1-hw21}
{Center for High Throughput Computing}, {\it Center for high throughput
  computing},  2006.

\bibitem{Callan:1977pt}
C.~G. Callan, Jr. and S.~R. Coleman, {\it {The Fate of the False Vacuum. 2.
  First Quantum Corrections}},  {\em Phys. Rev. D} {\bf 16} (1977) 1762--1768.

\bibitem{Vainshtein:1981wh}
A.~I. Vainshtein, V.~I. Zakharov, V.~A. Novikov, and M.~A. Shifman, {\it {ABC's
  of Instantons}},  {\em Sov. Phys. Usp.} {\bf 25} (1982) 195.

\bibitem{Andreassen:2016cvx}
A.~Andreassen, D.~Farhi, W.~Frost, and M.~D. Schwartz, {\it {Precision decay
  rate calculations in quantum field theory}},  {\em Phys. Rev. D} {\bf 95}
  (2017), no.~8 085011, [\href{http://arxiv.org/abs/1604.06090}{{\tt
  arXiv:1604.06090}}].

\bibitem{Andreassen:2017rzq}
A.~Andreassen, W.~Frost, and M.~D. Schwartz, {\it {Scale Invariant Instantons
  and the Complete Lifetime of the Standard Model}},  {\em Phys. Rev. D} {\bf
  97} (2018), no.~5 056006, [\href{http://arxiv.org/abs/1707.08124}{{\tt
  arXiv:1707.08124}}].

\bibitem{Ai:2019fri}
W.-Y. Ai, B.~Garbrecht, and C.~Tamarit, {\it {Functional methods for false
  vacuum decay in real time}},  {\em JHEP} {\bf 12} (2019) 095,
  [\href{http://arxiv.org/abs/1905.04236}{{\tt arXiv:1905.04236}}].

\bibitem{Ai:2018guc}
W.-Y. Ai, B.~Garbrecht, and P.~Millington, {\it {Radiative effects on false
  vacuum decay in Higgs-Yukawa theory}},  {\em Phys. Rev. D} {\bf 98} (2018),
  no.~7 076014, [\href{http://arxiv.org/abs/1807.03338}{{\tt
  arXiv:1807.03338}}].

\bibitem{Chigusa:2020jbn}
S.~Chigusa, T.~Moroi, and Y.~Shoji, {\it {Precise Calculation of the Decay Rate
  of False Vacuum with Multi-Field Bounce}},  {\em JHEP} {\bf 11} (2020) 006,
  [\href{http://arxiv.org/abs/2007.14124}{{\tt arXiv:2007.14124}}].

\bibitem{Appelquist:1981sf}
T.~Appelquist and U.~W. Heinz, {\it {Three-dimensional O(N) Theories at Large
  Distances}},  {\em Phys. Rev. D} {\bf 24} (1981) 2169.

\bibitem{Dilkes:1999uf}
F.~A. Dilkes and D.~G.~C. McKeon, {\it {Phase transitions in a scalar theory in
  (2+1)-dimensions}},  {\em Annals Phys.} {\bf 276} (1999) 1--11.

\bibitem{Weinberg:1987vp}
E.~J. Weinberg and A.-q. Wu, {\it {Understanding Complex Perturbative Effective
  Potentials}},  {\em Phys. Rev. D} {\bf 36} (1987) 2474.

\bibitem{Sher:1988mj}
M.~Sher, {\it {Electroweak Higgs Potentials and Vacuum Stability}},  {\em Phys.
  Rept.} {\bf 179} (1989) 273--418.

\end{thebibliography}
\end{document}